\documentclass{emulateapj}
\usepackage{amsmath}
\usepackage{enumitem}
\usepackage[flushleft]{threeparttable}
\usepackage{amsmath}
\usepackage{graphicx}
\usepackage{graphics}
\usepackage{longtable}
\usepackage{rotating}
\usepackage{natbib}
\usepackage{url}
\usepackage{hyperref}
\usepackage{rotating}
\DeclareMathSizes{11}{19}{13}{8}
\newcommand{\Mej}{M_{\text{ej}}}
\newcommand{\vej}{v_{\text{ej}}}

\slugcomment{In Prep. \today}

\shorttitle{Contamination in GW Optical Follow-up}
\shortauthors{Cowperthwaite \& Berger}

\begin{document}

\title{A Comprehensive Study of Detectability and Contamination in Deep Rapid Optical Searches for Gravitational Wave Counterparts}

\author{P. S. Cowperthwaite\altaffilmark{1,2} \& E. Berger\altaffilmark{1}} 

\altaffiltext{1}{Harvard-Smithsonian Center for Astrophysics, Cambridge, MA, 02138, USA, pcowpert@cfa.harvard.edu}
\altaffiltext{2}{NSF GRFP Fellow}

\begin{abstract}
The first direct detection of gravitational waves (GW) by the ground-based Advanced LIGO/Virgo interferometers is expected to occur within the next few years.  These interferometers are designed to detect the mergers of compact object binaries composed of neutron stars and/or black holes to a fiducial distance of $\sim 200$ Mpc and a localization region of $\sim 100$ deg$^2$.  To maximize the science gains from such GW detections it is essential to identify electromagnetic (EM) counterparts.  Among the wide range of proposed counterparts, the most promising is optical/IR emission powered by the radioactive decay of $r$-process elements synthesized in the neutron-rich merger ejecta -- a ``kilonova''.  Here we present detailed simulated observations that encompass a range of strategies for kilonova searches during GW follow-up.  We utilize these simulations to assess both the detectability of kilonovae and our ability to distinguish them from a wide range of contaminating transients in the large GW localization regions.  We find that if pre-existing deep template images for the GW localization region are available, then nightly observations to a depth of $i\approx 24$ mag and $z\approx 23$ mag are required to achieve a 95\% detection rate; observations that commence within $\sim 12$ hours of trigger will also capture the kilonova peak and provide stronger constraints on the ejecta properties.  We also find that kilonovae can be robustly separated from other known and hypothetical types of transients utilizing cuts on color ($i-z\gtrsim 0.3$ mag) and rise time ($t_{{\rm rise}}\lesssim 4$ days).  In the absence of a pre-existing template the observations must reach $\sim 1$ mag deeper to achieve the same kilonova detection rate, but robust rejection of contaminants can still be achieved.  Motivated by the results of our simulations we discuss the expected performance of current and future wide-field telescopes in achieving these observational goals, and find that prior to LSST the Dark Energy Camera (DECam) on the Blanco 4-m telescope and Hyper Suprime-Cam (HSC) on the Subaru 8-m telescope offer the best kilonova discovery potential.
\end{abstract}


\section{Introduction}
\label{sec:intro}
The era of gravitational wave (GW) astronomy is fast approaching with the Advanced Laser Interferometer Gravitational Wave Observatory (ALIGO, Abbott et al. 2009)  and advanced Virgo (AVirgo, Acernese et al. 2009) interferometers expected to make the first direct GW detections in the $2015-18$ time frame. These detectors will observe the GW signature resulting from the inspiral and merger of compact objects (e.g., neutron star-neutron star (NS-NS), solar mass black hole--black-hole (BH-BH), or NS-BH binaries). At design sensitivity, the ALIGO/AVIRGO network will be able to detect NS-NS mergers to an average distance of $\sim$ 200 Mpc (Abadie et al. 2010) and with a typical localization region of $\sim$ 100 deg$^2$ (Fairhurst 2009, Nissanke et al. 2011, Aasi et al. 2014). While this will be a remarkable achievement for both physics and astrophysics, the detection of an electromagnetic (EM) counterpart is essential to maximize the science obtained from a GW event. The identification of an EM counterpart will provide numerous benefits including: improved localization leading to host galaxy identification, determination of the distance and energy scales, identification of the progenitor local environment, and insight into the hydrodynamics of the merger (e.g., Sylvestre 2003, Stubbs 2008, Phinney 2009, Stamatikos et al. 2009, Fong et al. 2010, Metzger \& Berger 2012). Furthermore, the detection of an EM counterpart will assist in breaking modeling degeneracies to obtain more accurate measurements of binary properties from the GW signal (Hughes \& Holz 2003). 

Extensive theoretical and observational work has been performed to understand the potential EM emission from a merging compact object binary. The range of potential counterparts covers the entire EM spectrum. The most popular counterpart is a short-hard gamma-ray burst (SGRB; e.g., Paczynski 1986, Narayan et al. 1992, Berger 2014) powered by rapid accretion onto the compact remnant which leads to a strongly beamed relativistic jet. The evidence for a connection between SGRBs and compact object mergers includes the association with older stellar populations than for long GRBs (Berger et al. 2005, Bloom et al. 2006, Berger 2011, Fong et al. 2011, Fong et al. 2013), large physical offsets between the burst locations and their host galaxies indicative of progenitor kicks (Fong et al. 2010, Fong \& Berger 2013), weak association with the underlying stellar light of their hosts (Fong et al. 2010, Fong \& Berger 2013), and a lack of association with supernovae (Fox et al. 2005, Soderberg et al. 2006, Berger, Fong \& Chornock 2013, Berger 2014). The relativistic jet can also be observed in the x-ray, optical, and radio as non-thermal afterglow emission arises from interaction with the ambient medium. This emission is also affected by relativistic beaming but may still be seen off-axis as the jet decelerates and expands. In particular, for large off-axis angles $(\theta\gtrsim 2 \theta_j)$ the emission will be most pronounced in the radio on timescales of weeks to years (Nakar \& Piran 2011, Metzger \& Berger 2012).

The merger of a compact object binary involving at least one NS is also expected to produce mildly relativistic neutron-rich ejecta, with a typical mass of $M_{{\rm ej}} \sim 10^{-3}-10^{-1}\; M_{\odot}$ and a velocity of $\beta_{{\rm ej}} \sim 0.1-0.3$ (e.g., Rosswog et al. 1999, Rosswog 2005, Bauswein et al. 2013a). At later times, less neutron-rich outflows may arise from the resulting accretion disk, but the mass and composition will depend sensitively on the nature of the remnant (e.g., Metzger et al. 2009a, Dessart et al. 2009, Fern\'{a}ndez \& Metzger 2013, Fern{\'a}ndez et al. 2014). The dynamical ejecta is of particular interest when studying potential EM counterparts since it is expected to produce an isotropic quasi-thermal transient powered by the radioactive decay of heavy elements produced via {\em r}-process nucleosynthesis in the neutron-rich merger ejecta (Li \& Paczy\'{n}ski 1998, Rosswog 2005, Metzger et al. 2010, Grossman et al. 2014, Tanaka et al. 2014). The low ejecta mass and high ejecta velocity combined with the resulting {\em r}-process opacities ($\kappa \sim 10-100 \; \text{cm}^2 \text{ g}^{-1} \text{ for } \lambda \sim 0.3 - 3\; \mu\text{m}$, Kasen, Badnell, \& Barnes 2013) produces a significant suppression of bluer optical emission, resulting in a faint $(L \sim 10^{40}$ erg s$^{-1} \text{ at peak})$, rapid $(t \sim$ a few days), and extremely red transient with a peak at $\sim1.5$ microns (e.g., Kasen, Badnell, \& Barnes 2013, Barnes \& Kasen 2013, Tanaka \& Hotokezaka 2013), dubbed a ``kilonova." On longer timescales (e.g., $\sim$ years), the ejecta will decelerate via interaction with the ambient medium producing long-lasting isotropic emission in the radio (Nakar \& Piran 2011).

To investigate the detectability of kilonovae, and to asses potential contamination in deep high-cadence, wide-field searches for such sources, we carried out a set of observations using the Dark Energy Camera (DECam) mounted on the CTIO/Blanco 4m telescope (PI: Berger). This GW follow-up ``dry-run" covered an area typical for a GW error region ($\sim$ 70 deg$^2$), with two visits per night for five nights to a depth of $\approx$ 24 AB mag in both {\em i}- and {\em z}-band. This rich data set will serve as the framework to study the nature of foreground/background contamination and to test the effectiveness of various observing strategies and contaminant rejection strategies when performing follow-up observations of a GW event.

To help provide a framework for these observations, and to more broadly investigate the coupled issues of kilonova detectability and contaminant rejection, we present here a systematic study of the feasibility of kilonova detections and assessing the effects of a population of contaminant sources during follow-up of a GW trigger using wide-field optical telescopes. This study will guide the development of observational strategies for wide-field searches that maximize the likelihood of detecting a GW optical counterpart. The paper is organized as follows: In Section~\ref{sec:contaminants} we describe the kilonova models utilized in this work and develop the list of contaminating transients considered in this work. In Section~\ref{sec:phase} we  describe observational features that may be used to isolate kilonovae from the population of potential contaminants. In Section~\ref{sec:MCsims} we present detailed Monte Carlo simulations of observations to investigate the detectability of kilonovae and contaminant rejection for various observational strategies. In Section~\ref{sec:diff}, we address issues that arise from the possible lack of a deep pre-existing template. We then discuss the effects of modifications to the standard kilonovae models on our results in Section~\ref{sec:altkilo}. In Section~\ref{sec:disc}, we asses the expected performance of current and future wide-field telescopes in the framework of kilonova detectability and contaminant rejection and develop strategies for GW follow-up. Conclusions are presented in Section~\ref{sec:conc}.

Throughout this work all magnitudes are given in the AB system unless otherwise specified. For cosmological calculations, the Hubble constant is taken to be $H_0 = 67.77$ km s$^{-1}$ Mpc$^{-1}$ (Planck Collaboration et al. 2014). 

\section{Kilonovae and Contaminants}
\label{sec:contaminants}

To assess contamination from unrelated transients during a wide-field search for a GW optical counterpart, we consider various fast transients that may present kilonova-like behavior. primarily rapid evolution in their light curve, in rise and decline. We also consider Type Ia supernovae, as they are the most common SN type in large photometric surveys.

\subsection{Kilonovae}
\label{sec:kilo}
As discussed in Section~\ref{sec:intro}, the merger of NS-NS or NS-BH binaries is expected to produce a faint, rapidly-evolving optical/NIR transient powered by the radioactive decay of {\em r}-process elements formed in the ejecta. From a wide range of simulations it has been shown that the ejecta have a typical mass range of $\Mej \sim 10^{-3}-10^{-1} \; M_{\odot}$ with velocities of $\beta_{\text{ej}} \sim 0.1-0.3$, which depend on the initial mass asymmetry in the binary (Rosswog et al. 1999, Rosswog 2005, Bauswein et al. 2013a, Rosswog, Piran, \& Nakar 2013). The ejecta are expected to be neutron-rich, with typical electron fractions of $Y_e \lesssim 0.1$ (Metzger et al. 2010). 

A critical property for understanding kilonova emission is the opacity, as the ability of photons to diffuse through the ejecta determines the timescale, peak luminosity, and the spectral energy distribution of the transient. Metzger et al. (2010), using the models of Li \& Paczy\'{n}ski (1998, LP98 hereafter), found that a typical Fe-peak element opacity $(\kappa \sim 0.1 \; \text{cm}^2 \text{ g}^{-1} \text{ for } \lambda \sim 0.3 - 1\; \mu\text{m})$ leads to a peak time of $t_p \sim 1$ day, with a bolometric peak luminosity of $L_p \sim 10^{41} \text{ erg s}^{-1}$, and a photospheric temperature of $T_p \approx 10^4$ K (i.e., a peak in the UV). However, subsequent work by Kasen, Badnell, \& Barnes (2013) and Barnes \& Kasen (2013) found that {\em r}-process material have much larger opacities with $\kappa \sim 10-100 \; \text{cm}^2 \text{ g}^{-1} \text{ for } \lambda \sim 0.3 - 3\; \mu\text{m}$. This leads to a transient that is longer lived $(t_p \sim$ days), fainter $(L_p \sim 10^{40}$ erg s$^{-1})$, and much redder $(T_p \approx 3000$ K, e.g., a peak in the NIR). These results have been confirmed in simulations by other groups (e.g., Tanaka \& Hotokezaka 2013, Grossman et al. 2014, Tanaka et al. 2014).

 \begin{figure}[t!]
   \centering
	\includegraphics[width=1.1\columnwidth]{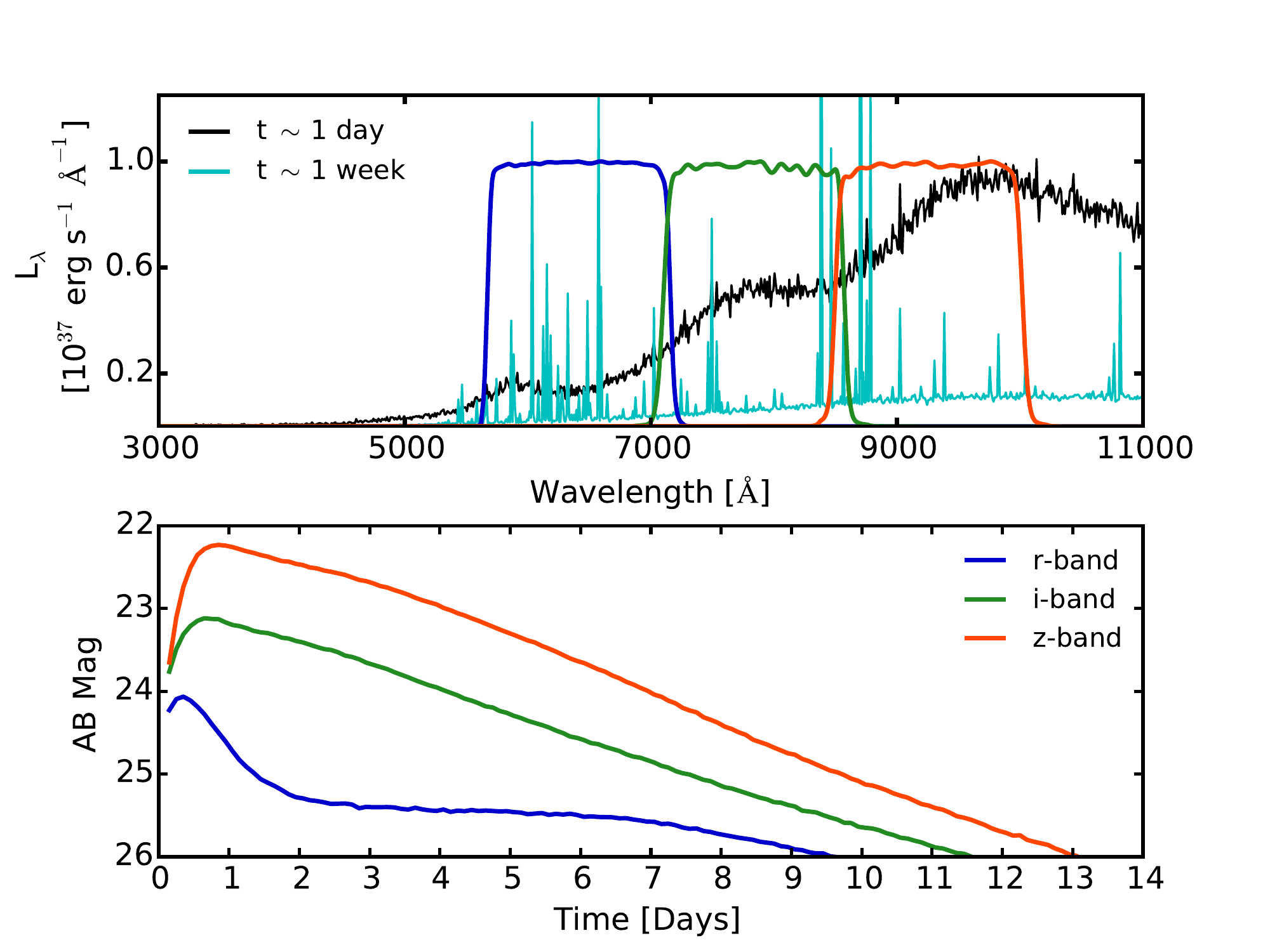}
   \caption{Top: Kilonova spectra from Barnes \& Kasen (2013) at $\sim$ 1 d and $\sim$ 1 week with ejecta properties of $\Mej =10^{-2}\;M_{\odot}$ and $\beta_{\text{ej}} = 0.2$. Bottom: Light curve for the same model parameters computed at a fiducial distance of 200 Mpc in the DECam {\em r}, {\em i}- and {\em z}-band filters. \vspace{10pt}}
   \label{fig:kilosample}
   \end{figure}

In this work, we construct kilonova light curves using the time-resolved spectra of Barnes \& Kasen (2013). They consider a grid of models with $\Mej = 10^{-3}\;, 10^{-2} \;, \text{ and } 10^{-1} \;M_{\odot}$ and with $\beta_{\text{ej}} = $ 0.1, 0.2, and 0.3. Unless otherwise noted we only consider the average ejecta properties of $\Mej =10^{-2}\;M_{\odot}$ and $\beta_{\text{ej}} = 0.2$ since we do not know the distribution of ejecta properties a priori. We compute light curves using the DECam filter transmission curves\footnote{\url{http://www.ctio.noao.edu/noao/sites/default/files/DECam/DECam_filters_transmission.txt}}\footnote{We note that the DECam transmission filters are fairly standard so these results are relevant for other instruments as well.} with the luminosity $L_{\phi}$, through a filter $\phi$, at a time t, being given by 
\begin{align}
\label{eq:LCint}
L_{\phi}(t) = \frac{\int d\lambda L_{\lambda}(t,\lambda) \phi(\lambda)}{\int d\lambda \phi(\lambda)}
\end{align}
\noindent where $L_{\lambda}(t,\lambda)$ is the model specific luminosity and $\phi(\lambda)$ is the filter transmission as a function of wavelength. A representative plot of the spectra at 1 day and 1 week, along with the resulting {\em r}-, {\em i}-, and {\em z}-band light curves are shown in Figure~\ref{fig:kilosample}.

\subsection{WD-NS/WD-BH Mergers}
\label{sec:WDmerge}
The tidal disruption of a WD by a NS or BH binary companion can produce an isotropic optical transient powered by the radioactive decay of ${}^{56}$Ni synthesized in the ejecta. The ejecta have an expected mass of $\Mej \approx 0.3-1.0 \; M_{\odot}$, velocity $\vej \approx (1-5) \times 10^4$ km s$^{-1}$, and $^{56}$Ni mass of $M_{\text{$^{56}$Ni}} \approx 10^{-4.5} - 10^{-2.5}\;M_{\odot}$ (Metzger 2012). These properties are primarily determined by the mass of the WD, with more massive progenitors producing higher ejecta mass, velocities, and $^{56}$Ni masses. The resulting light curves have a typical peak timescale of $t_p \sim $ week and a peak bolometric luminosity of $L_p \sim 10^{39} - 10^{41.5} \text{ erg s}^{-1}$.

The two scenarios we consider here are the merger of: (1) a $0.6\;M_{\odot}$ C--O WD with a $1.4\;M_{\odot}$ NS, and (2) a $1.2\;M_{\odot}$ O--Ne WD with a $3\;M_{\odot}$ BH. The models of Metzger (2012) only include bolometric light curves. To produce {\em i}- and {\em z}-band light curves we compute an effective temperature using the LP98 models with $\Mej \approx 0.5\;M_{\odot},\text{ and }\vej \approx 2.8\times10^4$ km s$^{-1}$ for scenario 1, and $\Mej \approx 1\;M_{\odot},\text{ and } \vej \approx 4.6\times10^4$ km s$^{-1}$ for scenario 2 (see Figure 7 of Metzger 2012). We integrate the resulting blackbody curve over the DECam filters using Equation~\ref{eq:LCint} and compare to the bolometric luminosity to generate bolometric corrections. For the WD-NS merger, we find $T_p \approx 6000$ K leading to $m_{\text{bol}} - m_{\text{i}} \approx -1.9$ mag and $m_{\text{bol}} - m_{\text{z}} \approx -2.2$ mag. The WD-BH merger calculation yields $T_p \approx 7850$ K leading to $m_{\text{bol}} - m_{\text{i}} \approx -2.2$ mag and $m_{\text{bol}} - m_{\text{z}} \approx -2.6$ mag. We assume that these corrections are constant in time. The bolometric corrections are applied to the light curves of Metzger (2012) to generate the light curves used in this analysis (see Figure~\ref{fig:LCcont}).

\subsection{White Dwarf Accretion Induced Collapse}
\label{sec:AIC}
The accretion of material onto an O--Ne WD near the Chandrasekhar mass can cause the degenerate star to undergo electron capture in its interior resulting in the formation of a NS, rather than a thermonuclear explosion. This process is known as accretion induced collapse (AIC). If the WD is rapidly rotating then a stable accretion disk will form. The disk will cool and become radiatively inefficient, driving powerful outflows with a characteristic ejecta mass of $\Mej \sim 10^{-2}\;M_{\odot}$ and velocity $\beta_{\text{ej}} \sim 0.1$. The outflowing material has an electron fraction of $Y_e \sim 0.5$ due to neutrino radiation from the proto-NS and synthesizes $\text{a few} \times 10^{-3}\; M_{\odot}$ of ${}^{56}$Ni powering an optical transient (Metzger et al. 2009b). Radiative transfer calculations by Darbha et al. (2010), motivated by the models of Metzger et al. (2009b), provided light curves of AIC events. They found that the resulting transient is rapidly evolving with the light curve peaking at $t_p \sim$ 1 day with a bolometric luminosity of $L_p \sim 2\times10^{41}$ erg s$^{-1}$.

We use the models of Dabhra et al. (2010) to compute \emph{i} and \emph{z}-band light curves using the same methodology as in Section~\ref{sec:WDmerge}. The effective blackbody temperature has an assumed fiducial value of $T_p \approx 5800$ K from the models of Dabhra et al. (2010, see their Figure 3). The resulting bolometric corrections are therefore $m_{\text{bol}} - m_{\text{i}} \approx -1.9$ mag and $m_{\text{bol}} - m_{\text{z}} \approx -2.2$ mag. We assume that these corrections are constant in time and apply them to the bolometric light curves to generate $i$- and $z$-band light curves (Figure 2).

\subsection{White Dwarf Double Detonation}
\label{sec:ELDD}
Detonation of a sufficiently large accreted He envelope $(M_{\text{He}} \sim 0.2\; M_{\odot})$ on the surface of a C--O WD can generate shock waves which then propagate to the core leading to the total detonation of the WD. In simulations performed by Sim et al. (2012), two possible detonation scenarios were considered. The first, edge-lit double detonation (ELDD), occurs when the initial detonation takes place at the surface of the WD. This produces a transient that is expected to reach a peak bolometric luminosity of $L_p \sim 10^{42}$ erg s$^{-1}$ at $t_p \sim 8$ d. The second scenario is the compression-shock double detonation (CSDD), where the core is compressed by the shock wave and C--O detonation occurs closer to the center of the WD. The transient is brighter than in the ELDD scenario with a peak bolometric luminosity of $L _p \sim 4\times10^{42}$ erg s$^{-1}$, and a correspondingly longer rise time ($t_p  \gtrsim 10$ d). In the CSDD scenario the transient is also longer-lived with no significant decline in brightness for $\sim 30$ d.

The slow evolution of the CSDD model light curve relative to kilonovae means that it can be easily removed from consideration as a contaminant on the basis of timescales alone. However, we still consider the ELDD scenario as the more rapid rise time and comparably faster decline means that it could serve as a potential contaminant. We make use of bolometric light curves from the simulations of Sim et al. (2012) to compute DECam {\em i}- and {\em z}-band light curves as in Section~\ref{sec:WDmerge}. The ELDD spectrum is approximated as a blackbody with a characteristic temperature of $T_p \approx 4800$ K using the model spectra of Sim et al. (2012, see their Figure 7). The bolometric corrections are found to be $m_{\text{bol}} - m_{\text{i}} \approx -1.7$  mag and $m_{\text{bol}} - m_{\text{z}} \approx -1.9$ mag. We assume that these corrections are constant in time and apply them to the bolometric light curves to generate $i$- and $z$-band light curves (Figure 2).

\subsection{Type .Ia Supernovae}
\label{sec:type.Ia}
Another possible WD-based transient is the detonation of an outer shell of helium on the surface of the WD. This can occur when the mass accretion rate from the donor is low enough that burning in the He shell becomes thermally unstable. The specific term ``type .Ia" was coined as a reference to He detonation in an AM Canum Venaticorum system (AM CVns) where the donor star is degenerate or semi-degenerate and accreting He rich material onto a C--O or O--Ne WD (Bildsten et al. 2007, Shen et al. 2010). The typical envelope mass is $M_{\text{env}} \sim \text{few}\times10^{-2}-10^{-1}\;M_{\odot}$ and the detonation ejects the bulk of the envelope at a typical velocity of $\vej \sim 10^4$ km s$^{-1}$. This leads to a transient with a typical peak time of $t_p \sim 2-10$ d and a peak bolometric luminosity of $L_p \sim 5\times10^{41} - 5\times 10^{42}$ erg s$^{-1}$ (Shen et al. 2010).

We produce type .Ia light curves from the time-resolved spectra computed from the models of Shen et al (2010). We consider the models $0.6+0.2\;M_{\odot}$, $0.6+0.3\;M_{\odot}$, $1.0+0.05\;M_{\odot}$, and $1.2+0.05\;M_{\odot}$, where the first number is the WD mass and the second number is the mass of the He envelope. We integrate the spectra over the DECam filter transmission curves as described in Section~\ref{sec:kilo} (Figure~\ref{fig:LCcont}).

\subsection{Pan-STARRS Fast Transients}
\label{sec:PSFast}
\begin{figure}[t!]
   \centering
	\includegraphics[trim= 0 0 0 240 ,clip,width=0.95\columnwidth]{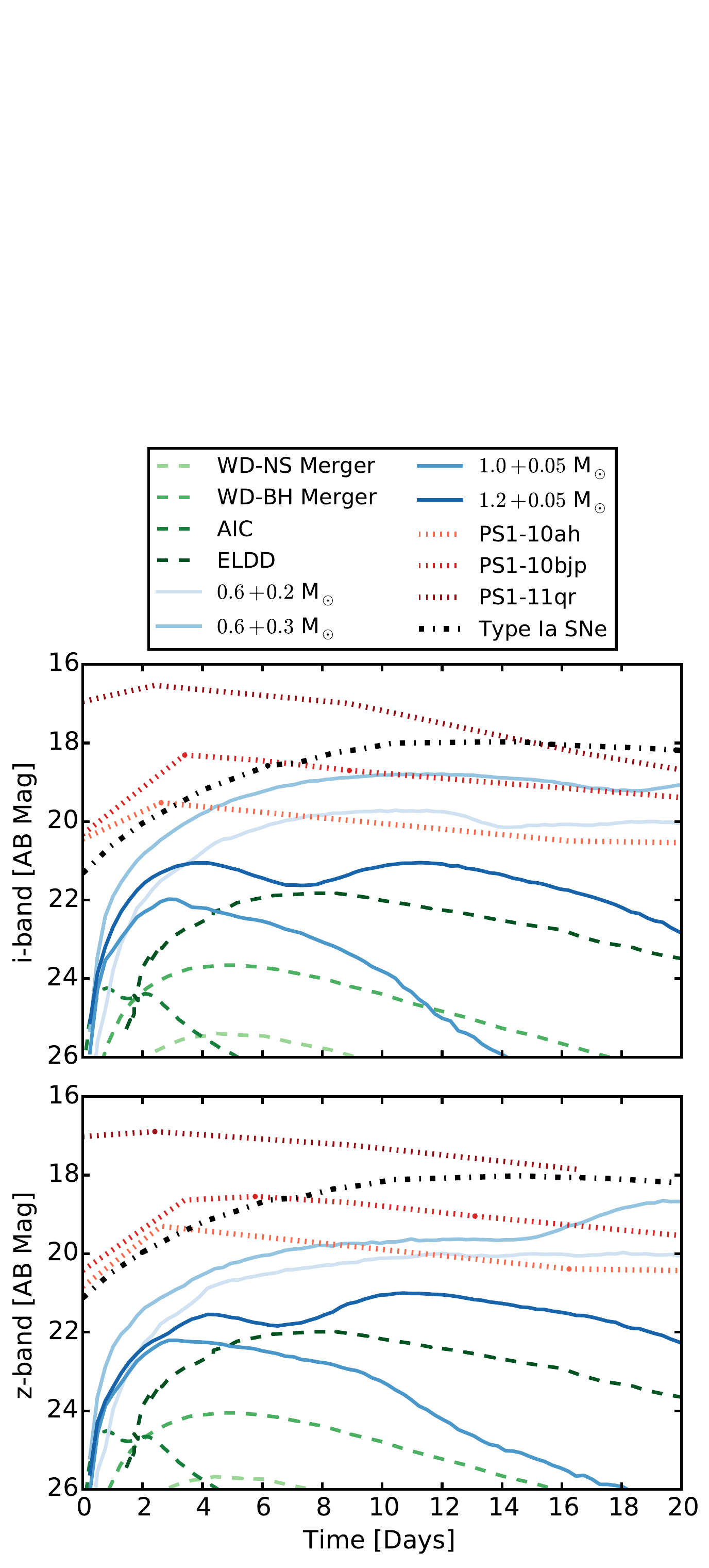}
   \caption{Top: {\em i}-band light curves of the contaminants considered in this work at a fiducial distance of 200 Mpc. Bottom: {\em z}-band light curves. See Sections~\ref{sec:WDmerge}~--~\ref{sec:typeIa} for details.}
   \label{fig:LCcont}
   \end{figure}

Drout et al. (2014) discovered ten rapidly evolving transients in the Pan-STARRS1 Medium Deep Survey. These events are characterized by a time above half-maximum of less than 12 days and have typical absolute magnitudes of $M \approx -16.5 \text{ to } -20$ mag. These objects have blue colors with {\em g} $-$ {\em r} $\lesssim$ --0.2 at peak (Drout et al. 2014). While the existing sample is small these events may end up being a significant contaminant in the fast cadence, wide-field GW searches investigated here. Therefore, we consider the {\em i}- and {\em z}-band light curves of three objects from Drout et al. (2014): PS1-10ah, PS1-10bjp, and PS1-11qr. They possess sufficient pre-peak data for our analysis while providing a representative range of timescales and luminosities for the sample (Figure~\ref{fig:LCcont}).

 \vspace{-5pt}
   
\subsection{Type Ia Supernovae}
\label{sec:typeIa}
Type Ia supernova result from the thermonuclear explosion of a C--O WD. While these SNe have timescales much longer than those of kilonovae and other transients considered in this section, their high volumetric rate and large luminosity mean that a deep optical search for kilonovae will detect large numbers of Type Ia SNe. This makes them a particularly likely contaminant in such searches. We focus our analysis on the event SN2011fe discovered in M101 by the Palomar Transit Factory (Nugent et al. 2011a) due to the high quality of available data. In addition to being one of the closest known Type Ia SNe ($\mu = 29.04\pm0.19$, Shappee \& Stanek 2011), SN2011fe was discovered a mere 12 hours after explosion (Nugent et al. 2011b) leading to coverage starting at --15 days relative to the {\em B}-band peak (Pereira et al. 2013). In this work, we use the 32 spectral observations compiled by Pereira et al. (2013) to produce DECam {\em i}- and {\em z}-band light curves as a function of redshift, including K-corrections, using the process described in Section~\ref{sec:kilo} (Figure~\ref{fig:LCcont}).

\begin{deluxetable*}{cccccc}[t!]
\tabletypesize{\footnotesize}
\tablecolumns{6} 
\tabcolsep0.05in\footnotesize
\tablewidth{0pt} 
\tablecaption{Expected Rates For Fast Transients}
\tablehead{
\colhead{Object}&  \colhead{$\mathcal{R}_{\rm vol}$}& \colhead{$z_{\rm max}$}  &  \colhead{$\mathcal{R}_{\rm area}$} &  \colhead{$N$}     &     \colhead{Reference}      \\
\colhead{} & \colhead{(Mpc$^{-3}$ yr$^{-1}$)} & \colhead{} & \colhead{(deg$^{-2}$ yr$^{-1}$)} &  \colhead{} & \colhead{} }
\startdata
 Type .Ia    &                  10$^{-6}$                   &  0.5        &  $0.7$            &         1.4  & Bildsten et al. 2007 \\
 WD-NS mergers  &                  10$^{-5}$                   &   0.06        & $2\times10^{-2}$              &         $3\times10^{-2}$  &     Thompson 2009     \\
 WD-BH mergers  &      10$^{-5}$                   &    0.06        &   $2\times10^{-2}$              &      $3\times10^{-2}$  &    Fryer et al.1999          \\
      AIC       &                  10$^{-6}$                   & 0.06        & $2\times10^{-3}$            &        $3\times10^{-3}$ &  Darbha et al. 2010   \\
      ELDD      &                  3$\times 10^{-7}$                   & 0.14        &    $6\times10^{-3}$           &      $1\times10^{-2}$ &  $\cdots$  \\
 Pan-STARRS fast &                  5$\times 10^{-6}$                   &     0.5        &   $3.5$            &     7 &   Drout et al. 2014  
 \enddata
\tablecomments{Expected rates for the various contaminants considered in Section~\ref{sec:contaminants}. $\mathcal{R}_{\rm area}$ is computed assuming an isotropic distribution of sources in a volume defined by the comoving volume at $z_{\rm max}$.  The column $N$ refers to the number of events expected during a search covering 100 deg$^2$ for 7 days. See \S~\ref{sec:rates} or details.}
\label{tab:rates}
\end{deluxetable*}

\vspace{-5pt}
\subsection{Stellar Flares}
\label{sec:flarestar}
Stellar flares, primarily from M dwarfs, are another potential contaminant in fast cadence wide-field optical searches (Becker et al. 2004, Kulkarni \& Rau 2006, Berger et al. 2013).  These flares have characteristic timescales of minutes to hours (e.g., Berger et al. 2013) and temperature of $T_p \sim 10^4$ K, leading to an appearance as fast and blue transients.  As a result, M dwarf flares will only appear as a significant contaminant in kilonova searches that rely on single-epoch detection, primarily in $g$- and $r$-band in which the kilonova timescale is shorter than in the redder bands.  We return to this point in Section~\ref{sec:disc}.  The sky-projected M dwarf flare rate at the depths relevant for kilonova searches is large, $\sim 10$ deg$^{-2}$ day$^{-1}$ (Kulkarni \& Rau 2006).  On the other hand, as was demonstrated by Berger et al. (2013), deep observations in the $i$- and $z$-band can easily reveal the underlying M dwarfs and hence reduce the potential for contamination from flares.

\subsection{AGN Variability}
\label{sec:agns}
We also consider short-duration optical variability from active galactic nuclei (AGNs). AGN typically show strong variability on month or year-long timescales, making AGN outbursts too long in duration to be confused with a kilonovae. However, recent work has detected rare cases of day-long  or shorter variability in the NIR. Genzel et al. (2003) observed an H-band flare in Sgr A$^{*}$ during which the source brightened by a factor of 5. This flare lasted $\sim$ 30 min, with a characteristic rise time of $\sim2-5$ min. Such short timescale flares can be ruled out as kilonova contaminants by requiring a detection on more than a single night. Flares with longer durations are also unlikely to be significant contaminants. Totani et al. (2005) performed {\em V}- and {\em i}-band observations of six AGN, selected on the basis of their variability, over a period of four days. They found typical variability of only $\sim1-5\%$ above their host galaxy luminosity. More recent work by Ruan et al. (2012) studied the variability of a sample of {\em Fermi} selected blazars. They found that these sources showed a typical rms variability of $\sim0.6$ mag on timescales of $\sim 10$ d. Therefore, AGN are unlikely to exhibit sufficiently large variability to be confused with a kilonova.

      \begin{figure*}[t!]
   \centering
	\includegraphics[trim= 0 0 0 250 ,clip,width=\textwidth]{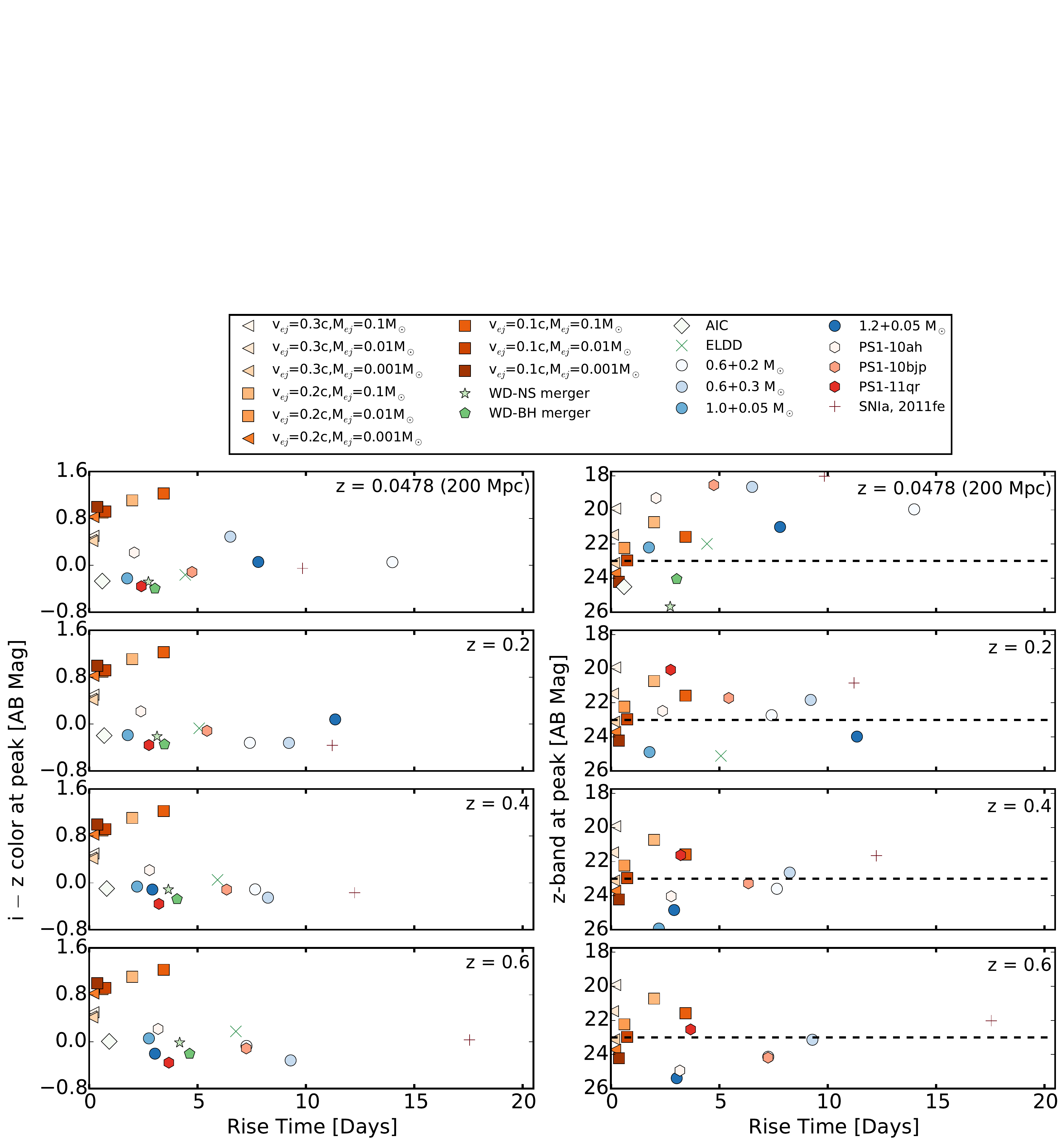}
   \caption{Time-color (left) and time-magnitude (right) slices of the time-color-magnitude phase-space at a selection of redshifts. Arrow markers denote upper limits on the rise time. Note that in all panels the population of kilonovae are kept at the fiducial distance of $\sim$ 200 Mpc while redshift corrections are applied to the contaminant population. Kilonovae are well separated from the contaminant population in both color and timescales. The delineation in magnitude is less apparent.}
   \label{fig:phase}
   \end{figure*}
   
\subsection{Volumetric Rates}
\label{sec:rates}
An additional important factor for understanding the impact of potential contaminants on fast cadence, wide-field searches is their volumetric rates. We note that the rates for these types of objects are generally poorly understood due to the lack of strong observational constraints; the notable exception are Type Ia SNe. The volumetric rates of potential contaminants are compiled in Table~\ref{tab:rates}. No values are provided in the literature for the ELDD volumetric rate so we assign a fiducial value of 1\% of the Type Ia SN rate, consistent with the other rapid transients in this study. The volumetric rate for Type Ia SNe is known to be a function of redshift, and we use an approximate average volumetric rate in redshift bins of $\Delta z \sim 0.25$ out to $z_{\text{max}} \sim 0.75$. (Li et al. 2012, Perrett et al. 2013,  Graur et al. 2014 and references therein). These rates are listed in Table~\ref{tab:rates_Iae}.

\begin{deluxetable}{cccc}
\tabletypesize{\footnotesize}
\tablecolumns{4} 
\tabcolsep0.05in\footnotesize
\tablewidth{0pt} 
\tablecaption{Expected Rates For Type Ia SNe}
\tablehead{
    \colhead{Redshift}     &  \colhead{$\mathcal{R}_{\rm vol}$}&  \colhead{$\mathcal{R}_{\rm area}$}  &     \colhead{$N$} \\
    \colhead{} & \colhead{(Mpc$^{-3}$ yr$^{-1}$)} &  \colhead{(deg$^{-2}$ yr$^{-1}$)} &  \colhead{} }
\startdata
0 -- 0.25 & $3\times10^{-5}$ & 3.5 & 21  \\
0.25 -- 0.50 & $4\times10^{-5}$ & 25 & 153  \\
0.5 -- 0.75 & $5\times10^{-5}$ & 65 & 394  
\enddata
\tablecomments{Expected rates for Type Ia SNe in redshift bins of $\Delta z = 0.25$ out to $z_{\rm max} = 0.75$. See \S~\ref{sec:rates}, Graur et al. (2014), and references therein.}
\label{tab:rates_Iae}
\end{deluxetable}

To estimate the expected number of events observed during a GW optical follow-up search we compute the number of sources per square degree per year. This is done by determining the maximum redshift to which the object can be observed to a limiting magnitude of $z \approx$ 23 mag (see Sections~\ref{sec:kilo} and~\ref{sec:MCsims_det}). This rate is then multiplied by the search area $(A \sim 100$ deg$^2$) and duration to get the number of expected sources for our observations. We compute the duration by assuming that a kilonova will be detected within approximately one day of the GW trigger. Therefore, we do not consider events that begin more than +2 d into the search. However, we do consider sources that begin up to twice their characteristic timescale ($\sim5$ d for rapid transients and $\sim 20$ d for Type Ia SNe) prior to the GW trigger, as they may remain visible when the search begins. The total duration is then 7 d for rapid transients and 22 d for Type Ia SNe. The number of expected sources is given in Tables~\ref{tab:rates} and~\ref{tab:rates_Iae}. Type Ia SNe dominate the number of expected events with $N \sim 568$ when summing across all redshift bins. Of the population of fast transients, only type .Ia and the Pan-STARRS fast transients are expected to have $N \gtrsim 1$. Despite the low expected occurrence rate for other fast transients, we still consider these sources in this analysis due to the large uncertainty in their rates.
   
\section{The Color-Magnitude-Timescale Phase-Space}
\label{sec:phase}
   
In this section we assess the possibility that any of the potential contaminants discussed in Section~\ref{sec:contaminants} could be confused  for a kilonova during a GW follow-up observation. We consider the maximum detection distance for a NS-NS binary merger to be 200 Mpc ({\em z} $\approx$ 0.05), while for the contaminating sources we utilize larger distances as appropriate. We establish a three-dimensional phase-space based on the peak {\em z}-band brightness, the {\em i} $-$ {\em z} color at {\em z}-band peak, and the rise time, defined as the time it takes the transient to rise one magnitude to peak in {\em z}-band. We compute the phase-space coordinates at 200 Mpc for all of the transients discussed in Section~\ref{sec:contaminants}. We then examine all of the potential contaminants at {\em z} $\sim$ 0.2, 0.4, and 0.6. In cases where time resolved spectra were available, the spectra were appropriately redshift-corrected in both time and wavelength. Sources modeled with a blackbody function had redshift corrections applied to the effective temperature. The rise time-color slice of this phase-space is shown in the left column of Figure~\ref{fig:phase} while the rise time-magnitude slice is shown in the right column of Figure~\ref{fig:phase}.

Investigating the first panel in the rise time-color space, we can see that the nine kilonovae models are well separated from the population of contaminating sources by virtue of their redder color. We also note that their rise times are generally shorter, with only the AIC and $1.0+0.05\;M_{\odot}$ type .Ia light curves exhibiting comparable timescales. The region occupied by kilonovae in the rise time-peak brightness phase-space is not as tightly constrained. They are still well separated by timescale but several of the contaminating sources have comparable peak brightnesses at 200 Mpc. As we consider more distant contaminant populations, we observe little to moderate reddening, with all of the contaminants remaining blue compared to the kilonovae population ($i-z\lesssim$ 0 mag to {\em z} $\sim$ 0.6). The $0.6+0.3\;M_{\odot}$ and $1.2+0.05\;M_{\odot}$ Type .Ia SNe models have $i-z\sim$ 0.2 mag, but their longer timescales $(t\sim\text{ week})$ keep them well separated from kilonovae. In addition, there is a systematic change in timescales, with all of the contaminants (except for the AIC events) shifting to a rise time of $\gtrsim  2$ d by {\em z} $\sim$ 0.6 due to time dilation. 
   
The inherent low luminosity of most fast transients means that they quickly drop below the sensitivity limit of a reasonable search (e.g., $m_z \approx 23$ AB mag). Only the Type Ia SNe, Pan-STARRS fast transients, and the most luminous type .Ia models are brighter than $\approx$ 23 AB mag by {\em z} $\sim$ 0.2, and by {\em z} $\sim$ 0.6 only the transient PS1-11qr and Type Ia SNe will still be detectable. 

Quantitatively, we can impose cuts of $t_{\text{rise}} \lesssim  5$ d, $i-z \gtrsim 0$ mag, and $m_z \lesssim 23$ mag to isolate kilonovae from contaminants. For sources at $z \lesssim 0.2$, this will isolate kilonovae from all contaminants except the Pan-STARRS fast transient PS1-10ah. These cuts remain effective for more distant sources as well. For example, at $z\sim0.6$, AIC, the $0.6+0.3\;M_{\odot}$ Type .Ia SNe, and PS1-10ah all exhibit $t_{\text{rise}} \lesssim  5$ d and $i-z \gtrsim 0$ mag. However, they no longer pass the brightness cut, with $m_z \gtrsim 23$ mag. Therefore, we conclude that given well-sampled light curves, we can use the rise time-color-brightness phase-space to separate kilonovae from the contaminant population of other rapidly evolving sources.

\section{Simulated Follow-Up Observations}
\label{sec:MCsims}
The analysis in the previous section was based on the idea that it is possible to obtain well-sampled light curves of kilonovae and other fast transients, but in reality the search cadence for a large area and the required depth will lead to at most a few light curve points. To explore the impact of realistic observations we utilize Monte Carlo (MC) simulations which model a set of GW follow-up observations parameterized by cadence, limiting magnitude, and start time following a GW trigger.  The goal of these simulations is two-fold. First, we study the feasibility of extracting useful light curve parameters for the kilonovae. This includes the actual detection of a kilonova, as well as its detection on the rise, which is important for establishing the ejecta parameters. Second, we study the impact of the contaminant population on our ability to identify kilonovae by utilizing our simulated observations to explore the rise time-color-magnitude phase-space established in Section~\ref{sec:phase}.

The detection of transient sources in wide fields necessitates difference imaging. In the simplest scenario a pre-existing image of the field can be used as a template to eliminate the non-varying sources. In the context of real-time GW follow-up, however, such template images may not be available until after the source fades away. To address this difficulty, we consider three scenarios:
\begin{enumerate}
\item The GW event is localized to a region with pre-existing template images. For example, the event occurs in a region previously observed to a similar or greater depth by the wide-field telescope being used for the follow-up.
\item Template images are obtained after the source has faded. For kilonovae follow-up, this requires observations at $\gtrsim10$ days after the GW trigger.
\item The first observation taken as part of a GW follow-up campaign is used as the template image.
\end{enumerate}

From an analysis standpoint, the first two scenarios are identical as both involve a template image which contains no flux from the transient source. This allows robust subtractions to be performed, as well as precise measurements of brightness and color. However, the second scenario does not allow for the real-time identification of the transient, which is critical for obtaining follow-up spectroscopy and NIR imaging. These scenarios are the primary focus of this section. 

The third scenario, using the first observation as a template, is more challenging. The potential for the presence of flux from the transient in the template image means that we can only measure a relative difference in brightness. When this difference is small compared to the noise in the image it becomes difficult to measure key quantities such as color or even claim that the source has been detected in difference imaging. We note that the second and third scenarios are not mutually exclusive. If the kilonova can not be identified in real-time under the third scenario, then a template image can still be obtained after the source has faded away and scenario 2 becomes applicable. These issues are addressed in Section~\ref{sec:diff}.

\subsection{Detectability of Kilonovae in Simulated Observations}
\label{sec:MCsims_det}
We first investigate the issue of kilonova detectability, as well as observation of the light curve during the rise to peak brightness. We explore four observing strategies utilizing $\sim$1 week of observations in both {\em i}- and {\em z}-band with the following cadences:
{\small \begin{enumerate}[leftmargin = 2.5cm]
\item[Strategy 1:] One visit per night for the duration of the search (i.e., N1, N2, N3, N4, N5, N6). 
\item[Strategy 2:]  One visit per night on the first two nights, followed by a visit every other night (i.e., N1, N2, N4, N6).
\item[Strategy 3:]  Two visits $\sim$ 3 hr apart on the first night, followed by one visit per night for the duration of the search (i.e., N1, N1, N2, N3, N4, N5, N6). 
\item[Strategy 4:]  Two visits $\sim$ 3 hr apart on the first night, followed by a visit every other night (i.e., N1, N1, N3, N5)
\end{enumerate}}
\noindent These strategies are chosen to probe a range of timescales while representing differing levels of time investment, and possible loss of coverage due to adverse weather during follow-up observations. 

\begin{figure*}[t!]
   \centering
	\includegraphics[width=0.7\textwidth]{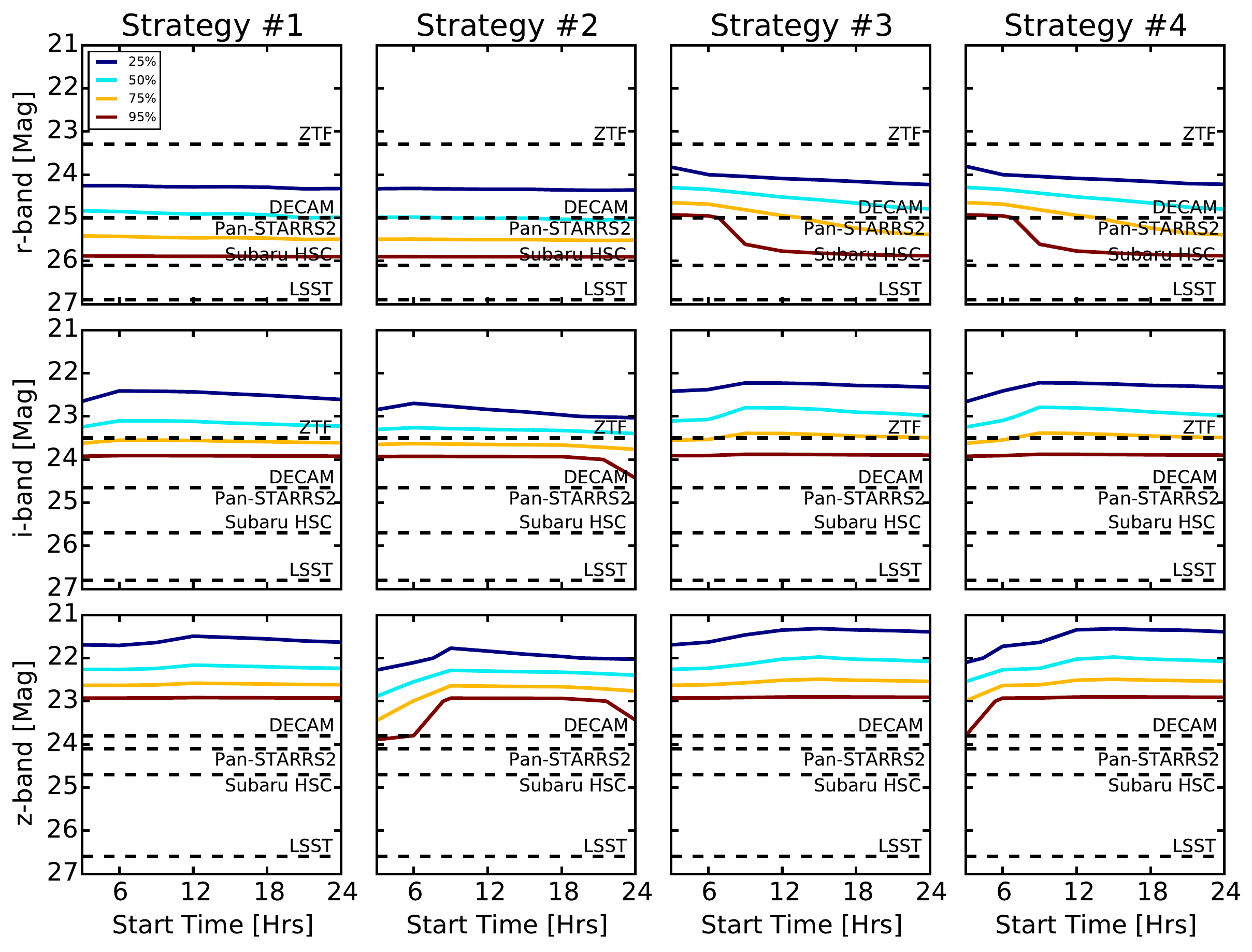}
   \caption{Depth versus start time grid in $r$ (top), $i$ (middle), and $z$ (bottom) for the four observing strategies. The contours indicate the fraction of sources detected in at least three epochs (25\% blue, 50\% cyan, 75\% yellow, 95\% red). The dashed lines represent the approximate 5$\sigma$ limiting magnitudes of several wide-field telescopes assuming the appropriate number of pointings necessary to cover a $\sim 100$ deg$^2$ error region with the required cadence. The choice of observing strategy does not drastically affect the fraction of sources detected at a given depth. We note that the early time drop in efficiency for {\em z}-band utilizing Strategy 2 is the result of the initial observations occurring before the light curve has brightened above the limiting magnitude.}
   \label{fig:det}
   \end{figure*}
   
\begin{deluxetable*}{ccccc|cccc}
\tabletypesize{\footnotesize}
\tablecolumns{9} 
\tabcolsep0.05in\footnotesize
\tablewidth{0pt} 
\tablecaption{Required 5$\sigma$ Limiting Magnitudes for Kilonova Detection}
\tablehead{\colhead{Filter} & \multicolumn{4}{c|}{With Pre-Existing Template} & \multicolumn{4}{c}{Without Pre-Existing Template} \\
\colhead{} & \colhead{Strategy \#1} & \colhead{Strategy \#2} & \colhead{Strategy \#3} & \multicolumn{1}{c|}{Strategy \#4} & \colhead{Strategy \#1} & \colhead{Strategy \#2} & \colhead{Strategy \#3} & \colhead{Strategy \#4}}
\startdata
$r$ & 24.9 (25.9) & 25.0 (25.9) &  24.5 (25.5) & 24.5 (25.5) & 25.0 (25.9) & 25.0 (25.9) & 25.0 (25.9) & 25.1 (25.9) \\
$i$ & 23.2 (23.9) & 23.3 (24.0) & 22.9 (23.9) & 22.9 (23.9) & 24.0 (24.9) & 24.3 (25.0) & 24.0 (24.9)& 24.3 (25.0) \\
$z$ & 22.2 (22.9) & 22.4 (23.2) & 22.1 (22.9) & 22.1 (23.0) & 23.1 (23.9) & 23.2 (24.0) & 23.1 (23.9) & 23.5 (24.2) 
 \enddata
\tablecomments{\,Limiting magnitudes $(5\sigma)$ required to detect 50\% (95\%) of kilonovae as a function of observing strategy and filter (Section~\ref{sec:MCsims_det}, Figure~\ref{fig:det}). Values are averaged over all start times. We compute the required depths for follow-up observations with and without a pre-existing template (Section~\ref{sec:diff}).}
\label{tab:det}
\end{deluxetable*}

For each observing strategy we establish a grid of start times following a GW trigger and a $5\sigma$ limiting magnitude. The range of grid values is $t_{\text{start}} = 3-24$ hr in steps of 3 hr and $m_{\text{lim}} = 21-27$ AB mag in steps of 1 AB mag. The range of start times represents the fact that a GW trigger may occur when the source is not readily visible in the night sky, while the range of limiting magnitudes covers the depths achievable by current and future wide-field instruments. At each point on this grid we conduct 50,000 simulated observations, with an observation being comprised of: (1) a kilonova with $M_{\text{ej}} = 10^{-2} \text{ M}_{\odot}$ and $\beta_{\text{ej}} = 0.2$ placed at a distance of up to 200 Mpc using a uniform volume distribution; (2) the light curve of the kilonova is computed for the chosen distance using the methodology outlined in Section~\ref{sec:kilo}; (3) the observation times are computed based on the chosen strategy; for example if the start time is 12 hours and strategy 1 is chosen, then observations will be conducted at 0.5 d, 1.5 d, 2.5 d, and so on; and (4) the source flux in $riz$-bands DECam filters is measured at the selected times using the light curves computed in step (2). If the kilonova brightness exceeds the search depth in at least three epochs, the event is flagged as a detection. If there is an observed increase in brightness between two epochs, the event is also flagged as ``rising."
   
    \begin{figure*}[t!]
   \centering
	\includegraphics[width=0.7\textwidth]{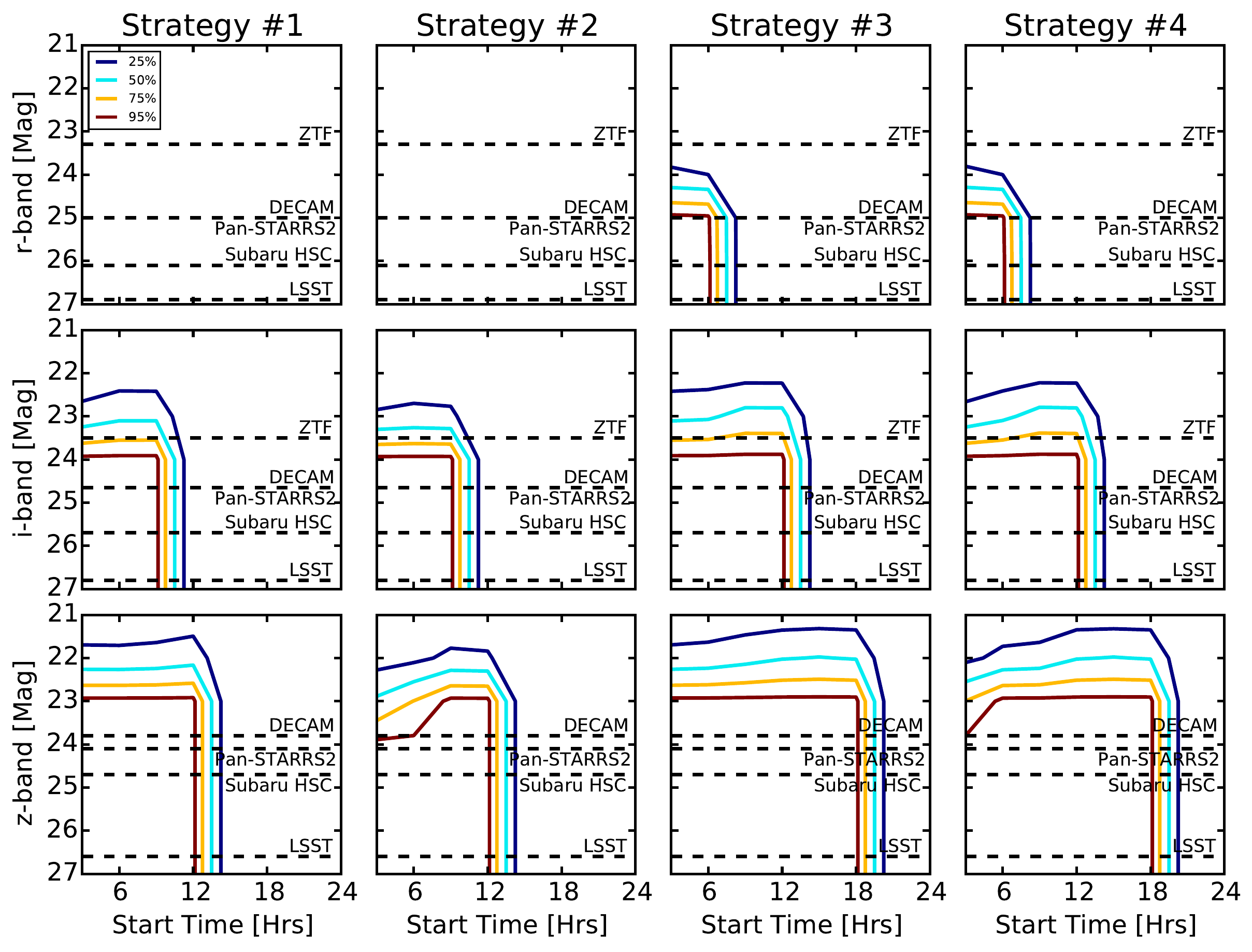}
   \caption{Same as Figure~\ref{fig:det}, except that the contours show the fraction of events for which a brightening is detected. The choice of observing strategy, start time, and filter have a strong impact on the detection of brightening.}
   \label{fig:rise}
   \end{figure*}

In Figure~\ref{fig:det} we plot contours of the fraction of events detected for each of the four observing strategies in the $riz$-bands, as a function of survey depth and start time. The dashed lines represent the approximate 5$\sigma$ limiting magnitudes of several wide-field telescopes assuming the appropriate number of pointings necessary to cover a $\sim 100$ deg$^2$ error region with the required cadence.

We find that the choice of observing strategy does not strongly impact the fraction of sources detected at a given depth. In particular, in {\em r}-band (Figure~\ref{fig:det}, top row) we find that for both Strategies 1 and 2, the typical depth required to detect 50\% of events is $\approx25$ mag, while a depth of $\approx26$ mag is required to detect 95\% of events. This challenge is slightly alleviated by Strategies 3 and 4, which involve two observations on the first night. In this case, an early start time $(\lesssim 6$ hr) allows 50\% (95\%) of events to be detected at a depth of $\approx24.3\,(25)$ mag. Consequently only Subaru HSC and LSST will be able to achieve the required depth for routine detections in $r$-band. The redder bands offer more favorable conditions, with the {\em i}-band 50\%(95\%) detection rate achieved at a depth of $\approx 23(24)$ mag across all four strategies. Lastly, we find that the {\em z}-band contours indicate that a depth of $\approx22.2\,(23)$ mag is required to achieve a 50\% (95\%) detection rate across all four strategies. In these cases, DECam, Pan-STARRS2, and Subaru HSC can all achieve the required depths for routine detections in $i$- and $z$-bands.  These values are summarized in Table~\ref{tab:det}.

We also find that for {\em z}-band the observing strategies utilizing longer cadences (i.e., Strategies 2 and 4) produce a noticeable drop in efficiency if the first observation occurs too early ($\lesssim 6$ hr). In this case, the light curve has not yet brightened above the survey depth, causing the initial observation to be a non-detection. The slow cadence after the initial night of observations (at $\gtrsim2$ days) is then insufficient to capture the requisite three detections before the light curve fades below the detection limit. Quantitatively, we find a loss of $\sim 1$ mag for Strategies 2 and 4 for starting times $\lesssim 6$ hr. This indicates that the strategies involving slower cadence after the initial night of observations are not well matched to kilonova timescales. 

While the particular choice of search strategy does not strongly impact the detection fraction at a given survey depth, the choice of strategy is important when trying to detect the {\em brightening} of a kilonova. We investigate this point in Figure~\ref{fig:rise}, which shows contours for the fraction of events that are seen to brighten between any two epochs. These events must also meet the detection criterion utilized in Figure~\ref{fig:det}. As in Figure~\ref{fig:det}, the contours are plotted as a function of start time and search depth for each of the observing strategies and filters. In {\em r}-band, utilizing Strategies 1 and 2, a brightening phase is not detected even to a limiting magnitude of 27 mag. This is because the {\em r}-band light curve peaks in $\lesssim 1$ day (Figure~\ref{fig:kilosample}) necessitating a rapid early time cadence to capture the event as it brightens. For Strategies 3 and 4, with two observations on the first night, brightening can be observed for 50\% (95\%) of events provided the observations reach a limiting magnitude of $\approx24.3\,(25)$ mag and begin within 8 (6) hr of the GW trigger.

The detection of the light curve rise is more favorable in {\em i}- and {\em z}-band. In general, the contours shown in Figure~\ref{fig:rise} are identical to those in Figure~\ref{fig:det}, but truncated as the start time becomes comparable to the peak time of the light curve. Therefore, brightening can be detected provided observations begin early enough and assuming the required depth is achieved. For {\em i}-band, with Strategies 1 and 2, we find that observations starting within 6 (8) hr of the GW trigger can detect brightening in 50\% (95\%) of events. With Strategies 3 and 4 the required start time for a detection rate of 50\% (95\%) extends out to 12 (14) hr. In {\em z}-band the observations can achieve detection rates of 50\% (95\%) provided they begin within 12 (15) hr of the GW trigger. The starting time can be extended to 18 (21) hr when Strategies 3 and 4 are utilized. We also note that Strategies 2 and 4 show the same early time ($\lesssim 6$ hr) loss of $\sim$ 1 mag as seen in Figure~\ref{fig:det}. The broader range of starting time allowed by strategies 3 and 4 highlight the advantage of rapid follow-up of GW triggers. However, these initial simulations consider any observed brightening as the detection of a rise. It may be the case that the actual change in brightness is not statistically significant, making a robust detection of a rise impossible. 

       \begin{figure*}[t!]
   \centering
	\includegraphics[width=0.7\textwidth]{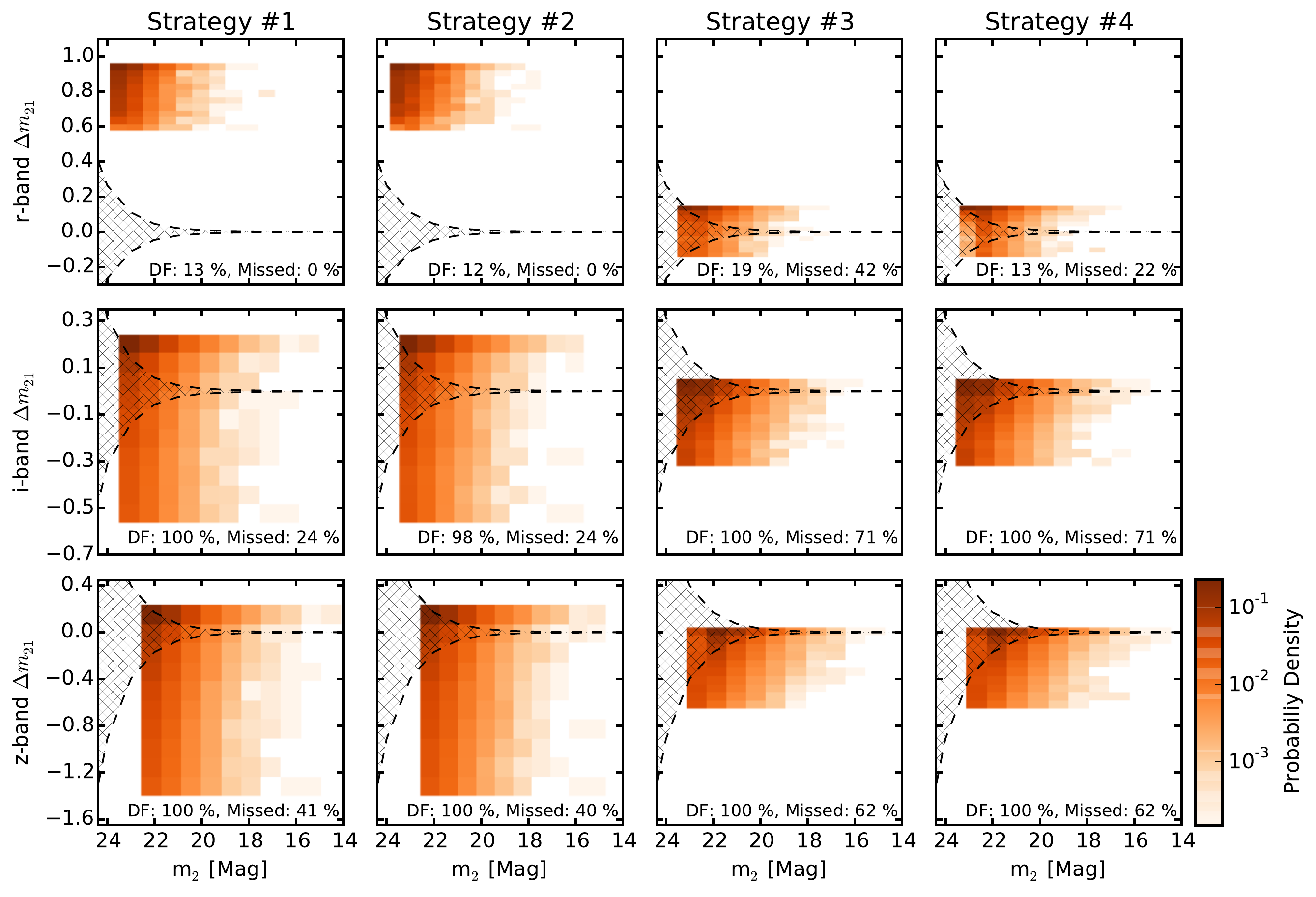}
   \caption{Probability density of kilonovae as a function of the change in brightness between the first and second observation $(\Delta m_21)$ versus the brightness in the second observation $(m_2)$ for each of the four observing strategies. The hatched region indicates the $2\sigma$ photometric error for DECam for an exposure time necessary to achieve the required depth. Points within this region will not exhibit a statistically significant change in magnitude for detection. The fraction of sources brighter than the detection limit in all three observations (DF) and the fraction of sources within the hatched region (missed) are indicated at the bottom of each panel. We find that the seemingly beneficial initial rapid cadence of Strategies 3 and 4 on a $\sim$3 hr baseline actually result in a large fraction of brightenings being undetectable.}
   \label{fig:delm2}
   \end{figure*}
    
We investigate the actual change in brightness relative to the instrument sensitivity using a modified version of our initial simulations. We no longer operate on the depth-start time grid, but instead consider a fiducial fixed depth of 24 AB mag in {\em r}- and {\em i}-band and 23 AB mag in {\em z}-band (as appropriate for achieving an acceptable detection fraction). The start time is randomly selected using a uniform distribution ranging from 3 to 24 hr. The same four observing strategies are employed and we run 50,000 iterations per strategy, utilizing the same procedure as before. We then compute the change in magnitude between the first two epochs ($\Delta m_{21} \equiv m_2 - m_1$), and between the first and third epochs ($\Delta m_{31} \equiv m_3 - m_1$). We employ the same detection criteria as in the previous simulations, requiring the source to be detected in at least three epochs. We consider sources that are observed to either rise or decline.
   
\begin{figure*}[t!]
   \centering
	\includegraphics[width=0.7\textwidth]{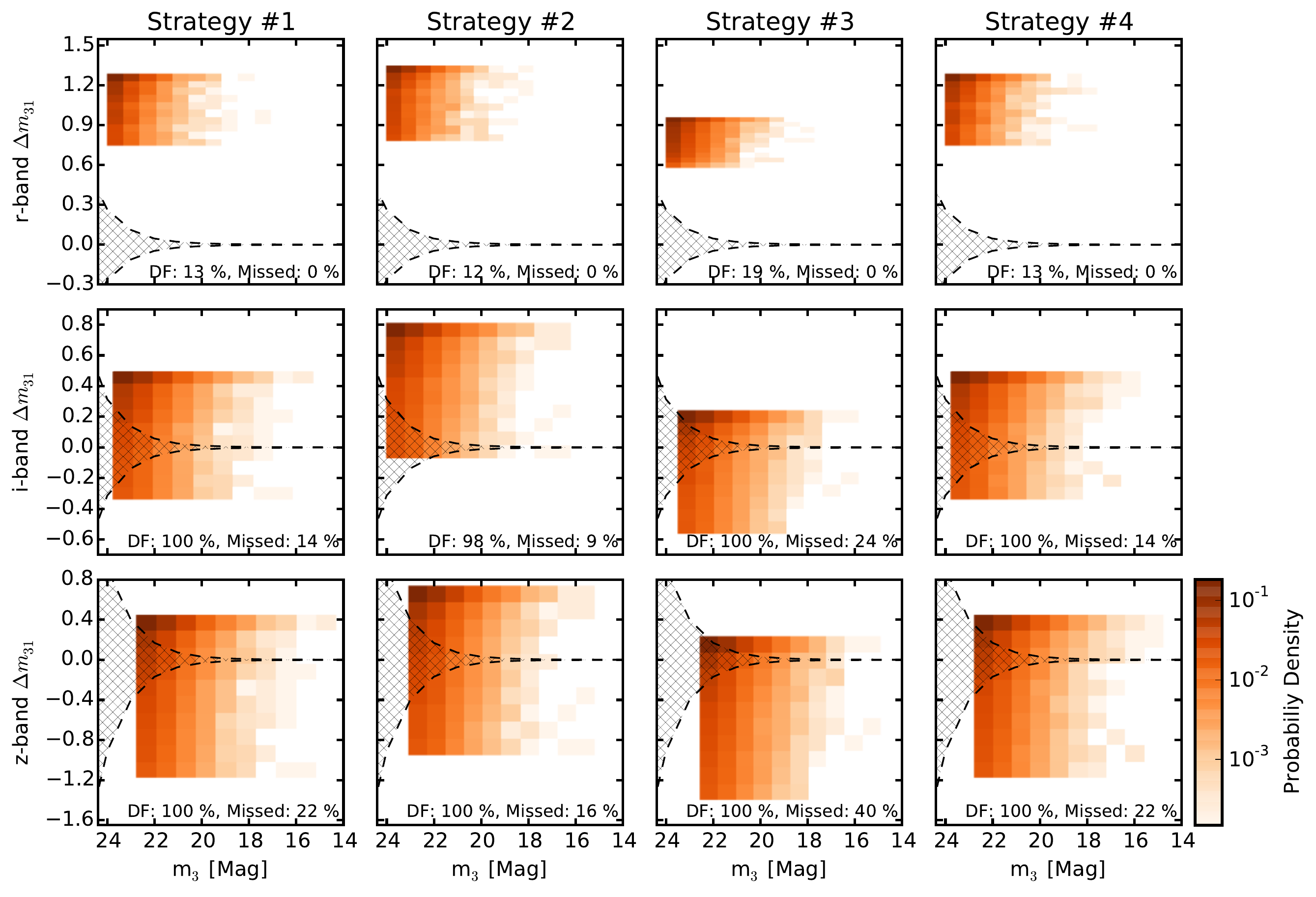}
   \caption{Same as Figure~\ref{fig:delm2}, except we plot the change in brightness between the first and third epoch $(\Delta m_{31})$ versus the brightness in the third epoch $(m_3)$. We find that the increased time between observations allows a much larger fraction of sources to exhibit a detectable change in brightness.}
   \label{fig:delm3}
   \end{figure*}

In Figure~\ref{fig:delm2} we plot two-dimensional histograms of $\Delta m_{21}$ versus $m_2$ for each of our four observing strategies and each of the three filters. The resulting histogram is color-coded by the probability of a kilonova falling in that bin, with darker colors indicating a higher probability. We also compute the $2\sigma$ photometric error for an exposure time necessary to reach our chosen target depth in each filter using the DECam exposure calculator\footnote{\url{http://www.ctio.noao.edu/noao/content/Exposure-Time-Calculator-ETC-0}} (hatched region). Sources that exhibit a change in magnitude that is smaller than this photometric error will not produce a statistically significant detection of a change in brightness. Figure~\ref{fig:delm3} follows a similar construction, showing the two-dimensional histograms for $\Delta m_{31}$ versus $m_3$.

Using these results we determine the fraction of sources that do not exhibit a sufficient change in brightness to be identified as a transient after the second or third observation. We begin by considering the first and second columns of Figures~\ref{fig:delm2} and~\ref{fig:delm3} (Strategies 1 and 2). Recall that for these strategies the time between the first and second epoch is $\sim$1 day. In both of these cases, the rapid evolution of the {\em r}-band light curve means that there is always a detectable change in brightness, but the source will always be seen in decline. Furthermore, it is still the case that given the overall faintness of kilonovae in {\em r}-band, only $\sim$13\% of sources will be detected at $r\approx24$ mag.

In {\em i}- and {\em z}-band we find that $(\sim 20-40\%)$ of sources are missed due to an insufficient change in brightness. For Strategies 3 and 4 where the time between the first two observations is only $\sim$3 hr over half of the sources $(60-70\%)$ fail to exhibit a statistically detectable change in brightness. This indicates that the intra-night observations have a time baseline that is too short for detecting a change in brightness even for the fast kilonova timescale.
   
Figure~\ref{fig:delm3} shows observations that are separated by a longer timescale $(m_3 - m_1)$: For Strategies 1 and 4, the first and third epoch are separated by 2 days, while the same visits are separated by 4 days for Strategy 2, and only 1 day for Strategy 3. For all four strategies we find that {\em r}-band exhibits a sufficient change in brightness, with $\Delta m_{31}$ well outside of the hatched region, but these sources are always observed to decline. As before, only $\sim10-20$\% of the sources are actually detectable at $r\approx24$ mag. In both {\em i}- and {\em z}-band we find that it is possible to detect a change of brightness in a much higher fraction of sources compared to $\Delta m_{21}$. Strategies 1 and 4 both show that $\sim 15-20\%$ of sources will not show a sufficient change in brightness. The highest fraction of missed sources is $\sim 25-40\%$ for Strategy 3. This is expected given the short separation between epochs. Strategy 2, which has the longest separation between epochs, shows only $\sim 10-15 \%$ of sources remaining undetectable. Furthermore, while sources in $r$-band are only observed to decline, in $i$- and $z$-band the rise can also be measured without requiring a $\sim$3 hr baseline. This further highlights the importance of having observations that are sufficiently separated in time to probe the kilonova light curve across a wide range of its evolution.
   
The rapid decline of $\sim0.6-1.3$ mag in {\em r}-band that is evident in Figures~\ref{fig:delm2} and~\ref{fig:delm3} hints at a possible detection strategy. Namely, the GW error region can be observed with a cadence of $\sim$1 d, with rapidly declining sources flagged. However, a depth of $\approx$26 mag is required for a high detection fraction even if we require only 2 detections (to separate kilonovae from stellar flares), so covering the large GW error region can only be achieved with LSST.
   
These detectability simulations indicate that using intra-night observations (Strategies 3 and 4) to detect the rise of the light curve is not an efficient use of time for follow-up observations since kilonovae generally do not exhibit a statistically significant change in brightness on this timescale. As shown in Figure~\ref{fig:det}, the choice of cadence does not drastically impact the likelihood of detection. Therefore, the most efficient solution, from a detectability standpoint, is Strategy 1. It provides a sufficient cadence to effectively detect kilonovae while maximizing the amount of area that can be covered in a single night. Observations should be carried out in {\em i}- and {\em z}-band as the depths required to detect 95\% of sources $(\approx 24,23$ AB mag, respectively), are obtainable by existing and future facilities.

\subsection{Simulating the Effects of Contamination}
\label{sec:MCsims_cont}
   
We now use our simulated observations to explore the contaminant population in the framework of the rise time-color-magnitude phase-space outlined in Section~\ref{sec:phase}. We accomplish this by repeating the simulations for each of the contaminants considered in Section~\ref{sec:contaminants}. We then utilize the simulated observations to compute analogues to the rise time-color-magnitude phase-space for both kilonovae and the contaminant population. In this manner we can produce a version of the phase-space that takes into account a distribution of contaminants with realistically sampled light curves given the choice of cadence, depth, and start time appropriate to wide-field follow-up observations of a GW trigger.

       \begin{figure*}[t!]
   \centering
	\includegraphics[width=0.7\textwidth]{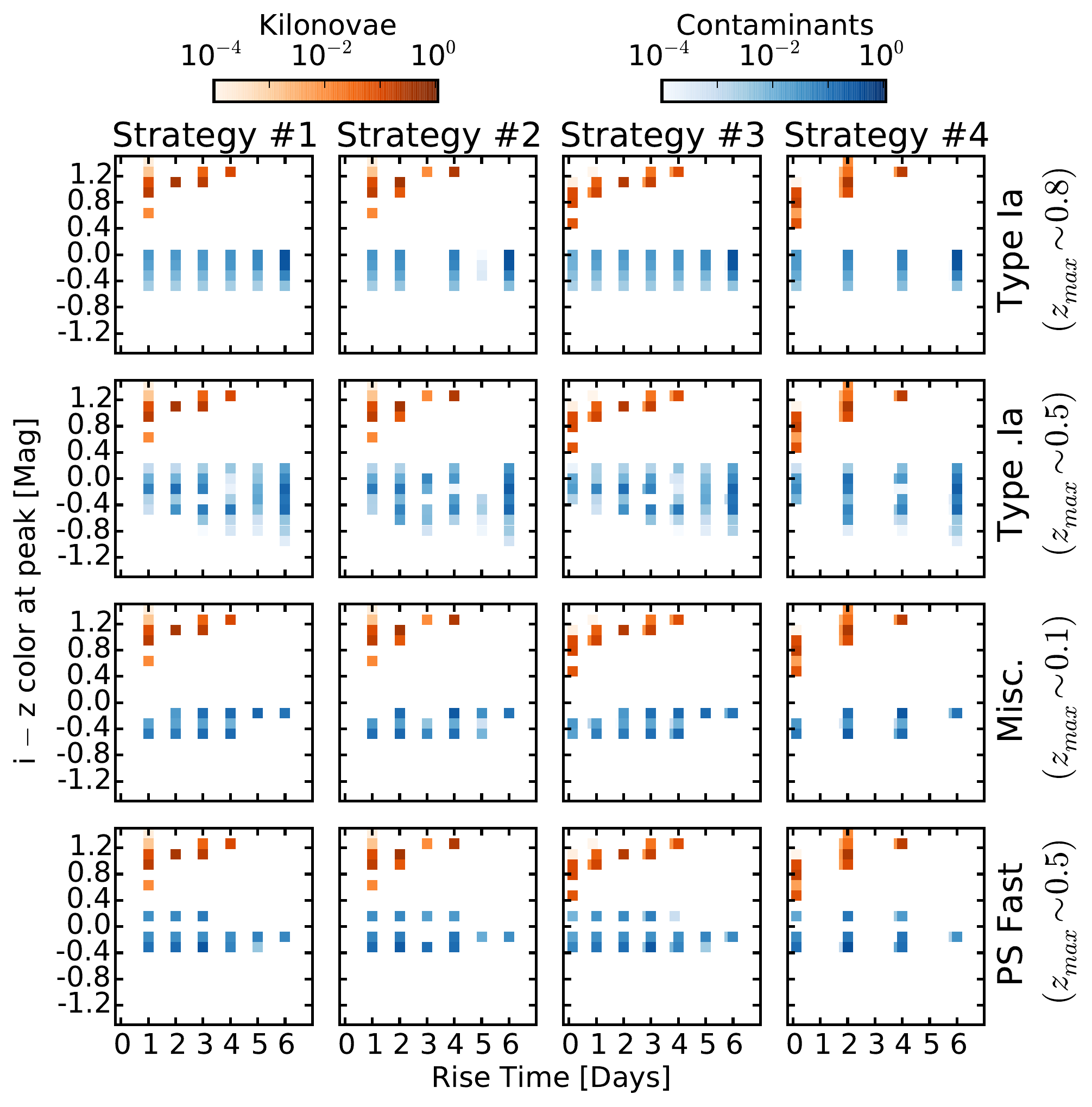}
   \caption{Simulated rise time-color phase-space for each of our four observing strategies. In each panel, we consider the full grid of kilonova models to a distance of 200 Mpc (orange). Each row highlights a different subset of the contaminant population over the comoving volume to $z_{\text{max}}$ (blue). The offset bins for Strategies 3 and 4 are due to the availability of intra-night observations on the first night. We find that the kilonovae are much redder than any potential contaminant, but the separation is less well-defined when compared to the light curves used in Section~\ref{sec:phase}.}
   \label{fig:mcphasecol}
   \end{figure*}
   
    \begin{figure*}[t!]
   \centering
	\includegraphics[width=0.7\textwidth]{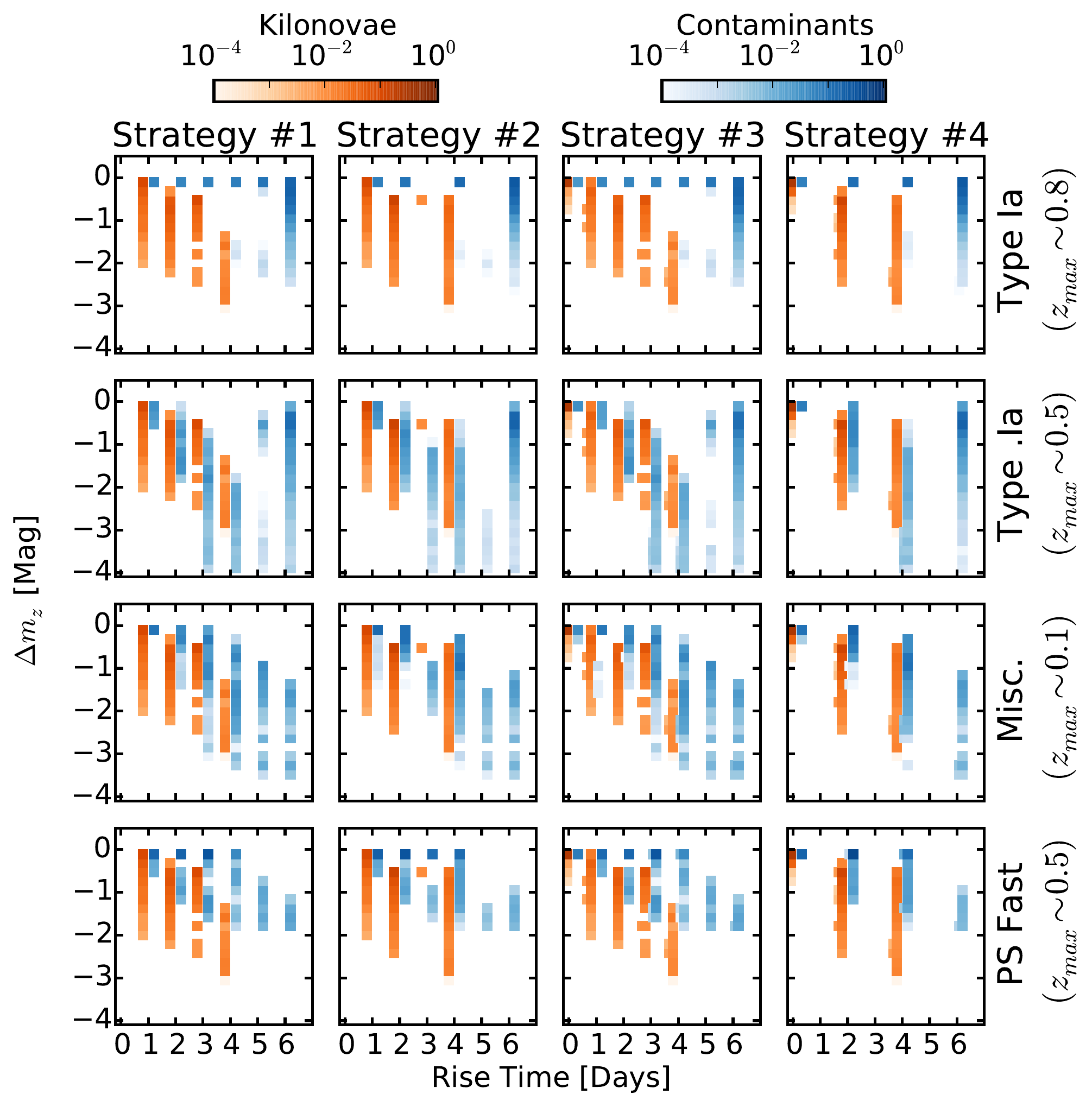}
   \caption{Same as Figure~\ref{fig:mcphasecol}, but showing the simulated rise time-$\Delta m_z$ slice of the phase-space. In this slice, there is much more confusion between the kilonovae and contaminant populations. With the exception of Type Ia SNe, there is significant overlap between kilonovae and the contaminant populations. This is the result of considering rapidly evolving sources for our contaminants, and it highlights the importance of color information in the selection of kilonovae candidates.}
   \label{fig:mcphasedm}
   \end{figure*}

We begin by employing the same four observing strategies as in the previous simulations. We again consider observations at a fixed depth of $\approx$ 24 AB mag in {\em i}-band and $\approx$ 23 AB mag in {\em z}-band, with the observations beginning 3 to 24 hr after the GW trigger. The simulations include 50,000 iterations, each consisting of: (1) A contaminant placed at a random, volume-weighted redshift. We consider contaminants at much larger distances than 200 Mpc to account for the fact that many contaminants are more luminous than kilonovae; the choice of $z_{\text{max}}$ is indicated in Figure~\ref{fig:mcphasecol}. (2) The contaminant light curve is randomly placed out of phase with respect to the start time of observations. We consider contaminants that occur approximately twice their characteristic timescale prior (5--20 days) and up to 2 days after the start of observations. The latter value is motivated by the results of the previous section in which we showed that kilonovae should be detectable within 1--2 days, and therefore, sources that first appear more than 2 days after the start of the observations can be rejected. Once the phase offset is determined the observation times are computed and the source brightness in {\em i}- and {\em z}-band is computed at each epoch. (3) The observed peak of the {\em z}-band light curve is measured. (4) The {\em i} -- {\em z} color is computed at the epoch where {\em z}-band attains its maximum value. (5) The rise time is defined as the time between the {\em z}-band peak and the first observation in which the source is detected. In cases where the source is only observed in decline, then the decline time is defined as the time between the first observation and the last observation in which the source is detected. 
   
In Figure~\ref{fig:mcphasecol} we plot the rise time-color slice of the three-dimensional phase-space for each observing strategy. In addition to the nine kilonovae models, we consider four groups of sources with each given an equal contribution. The first group of contaminants is Type Ia SNe (Figure~\ref{fig:mcphasecol}, first row). The second group is comprised of the four type .Ia SNe models (Figure~\ref{fig:mcphasecol}, second row). We form the third group from all of the low luminosity, fast transients (i.e., AIC, WD-NS/BH mergers, and ELDD; Figure~\ref{fig:mcphasecol}, third row). Lastly, the fourth group of contaminants contains the three Pan-STARRS fast transients (Figure~\ref{fig:mcphasecol}, fourth row).

    \begin{figure*}[t!]
   \centering
	\includegraphics[width=0.7\textwidth]{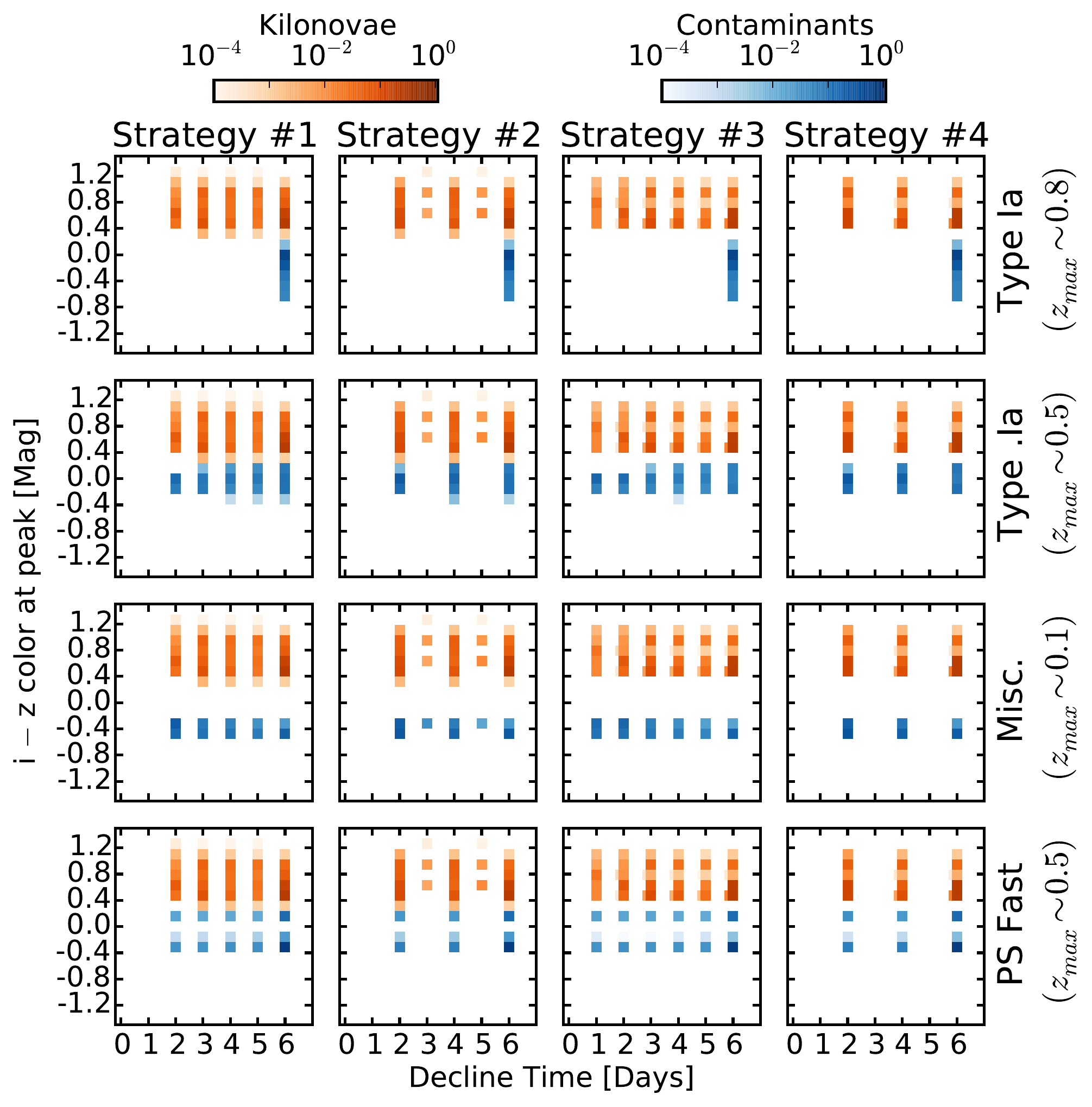}
   \caption{The decline time-color slice of the decline time-color-magnitude phase-space. The plot is constructed using the same methodology as in Figures~\ref{fig:mcphasecol} and~\ref{fig:mcphasedm}. We note that, as before, kilonovae show redder $i-z$ colors than the contaminant population. However, there is no strong separation in decline time.}
   \label{fig:deccol}
   \end{figure*}
   
To analyze the results, we first bin sources by their rise times, which fall in discrete values due to the cadence of the observations. For each rise time bin, we then bin the source populations by $i-z$ color in equally spaced bins from --1.5 to +1.5 mag. For each bin, we compute the fraction of detected sources in that bin relative to the total number of detected sources.

As we showed in Section~\ref{sec:phase}, kilonovae stand out from the contaminant population primarily on the basis of their redder colors, even when luminous contaminants are considered to higher redshifts (e.g., Type Ia SNe, Pan-STARRS fast transients). For all choices of observing strategy there is essentially no overlap in $i-z$ color between the kilonova population and any of the contaminant populations. We note that across all four observing strategies $\sim$12\% of Type Ia SNe, $\sim7\%$ of type .Ia, and $\sim8\%$ of Pan-STARRS fast transients exhibit $i-z\gtrsim 0$, while none have $i-z\gtrsim0.4$ mag. However, all kilonovae exhibit $i-z\gtrsim0.4$ mag. These numbers are relative to the total number of events observed during the rise and do not include sources that are observed only in decline (see below). Consequently, $i-z$ color remains a strong discriminant for separating kilonovae from the contaminant population. 
   
So far we have neglected the potential effect of extinction in reddening the $i-z$ colors of the contaminant sources. If the extinction is sufficiently large an intrinsically blue contaminant may be reddened into the regime of kilonova $i-z$ colors. However, this extinction will also cause the source to appear dimmer and may therefore push it below the detection limit. To quantify the effects of extinction we use the Milky Way extinction curve with $R_V = 3.1$. Figure~\ref{fig:mcphasecol} indicates that a reddening of $E(i-z) \approx 1$ mag is required to shift the contaminant population into the region of phase-space occupied by kilonovae. This will result in an extinction of $A_V \approx 6.3$ mag, $A_i \approx 3.9$ mag and $A_z \approx 2.9$ mag.  However, in our simulation nearly all ($\gtrsim 97\%$) of the contaminants which pass the detection cuts have $m_z \gtrsim 20$ mag meaning that an extinction of 2.9 mag will dim these sources below our {\em z}-band detection limit of $\approx$ 23 mag. Thus, reddening due to dust extinction does not seem to be a concern.

We also consider the possibility of separating kilonovae by their rise time. In Figure~\ref{fig:phase}, the entire set of kilonovae models showed generally shorter timescales relative to the contaminant population with rise times of $\lesssim4$ days. However, when the contaminants are placed out of phase with the GW trigger, and the observing cadence is taken into account, we find contaminants that present an apparent shorter rise time. This effect is the most prominent in the Pan-STARRS fast transients and miscellaneous fast transients for which $\sim55-60\%$ and $\sim65-75\%$ of detected sources show a rise time of $\lesssim$ 4 days, respectively. We also find that $\sim 35-45\%$ of detected type .Ia show a rise time that is $\lesssim$ 4 days. Lastly, for Type Ia SNe, $\sim20-25\%$ of detected sources will present a rise time of $\lesssim$ 4 days. Thus, rise time alone is not a strong discriminant for separating kilonovae from the contaminant population.

    \begin{figure*}[t!]
   \centering
	\includegraphics[width=0.7\textwidth]{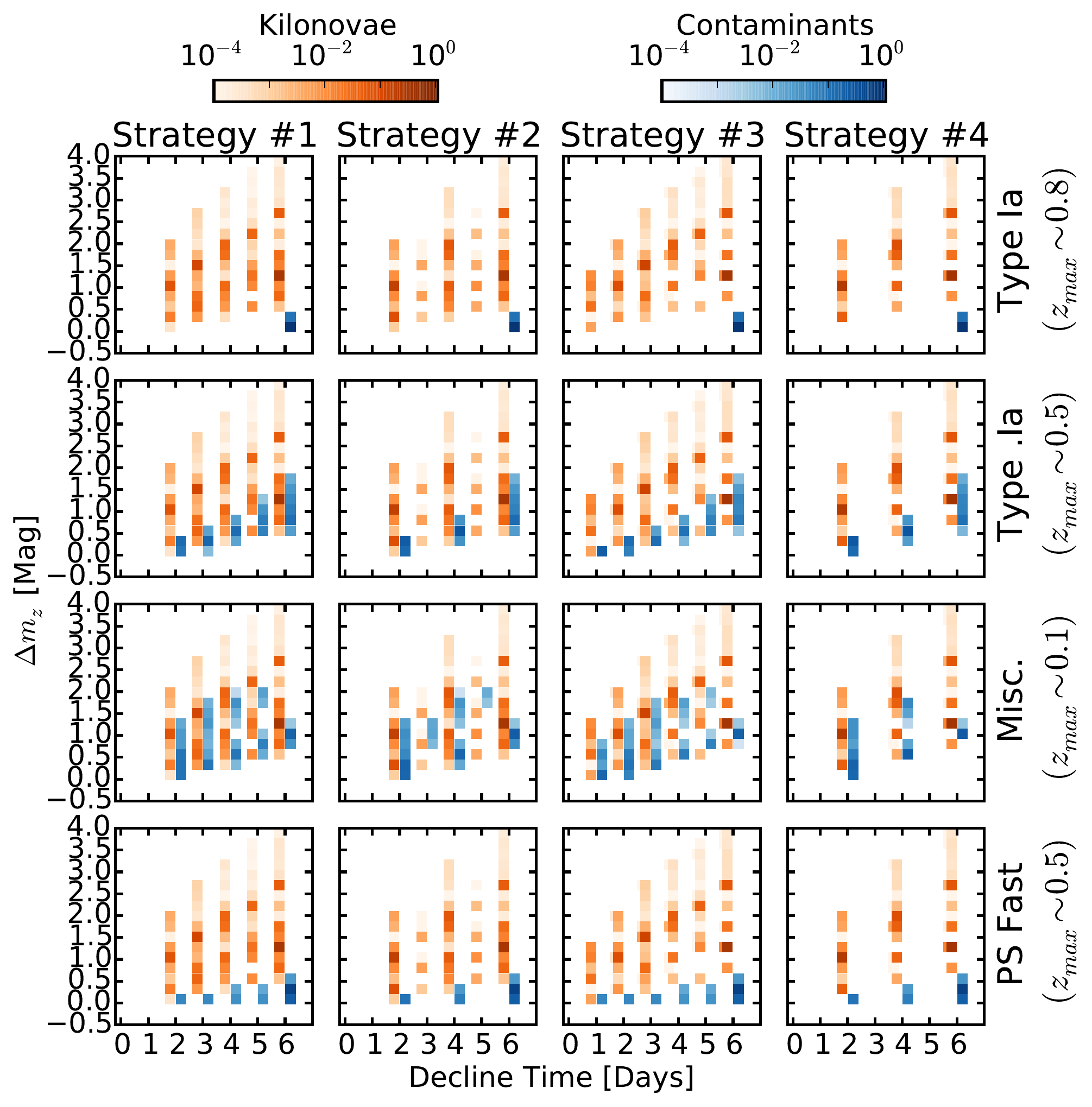}
   \caption{The decline time-$\Delta m_z$ slice of the decline time-color-magnitude phase-space. The slow decline of Type Ia SNe and the Pan-STARRS fast transients separate them from the rapidly evolving kilonovae. However, the type .Ia and the miscellaneous fast transients show changes in brightness comparable to those of kilonovae.}
   \label{fig:decdm}
   \end{figure*}
   
In Section~\ref{sec:phase}, we considered the peak {\em z}-band magnitude as the third data point in our phase-space. However, in a magnitude limited search, the population of contaminants will be dominated by events at the furthest distance to which they can be detected. Here we instead track the evolution of the source brightness by computing the difference in {\em z}-band magnitude between the source at the observed peak ($m_{z,{\rm peak}}$) and the first observation in which it is detected ($m_{z,0}$), denoted as $\Delta m_z \equiv m_{z,{\rm peak}} - m_{z,0}$. We show the rise time-$\Delta m_z$ slice of the three-dimensional phase-space in Figure~\ref{fig:mcphasedm}. The basic layout of the figure is identical to that of Figure~\ref{fig:mcphasecol}, but sources are offset by $\pm$5 hours to improve visibility.

We find that kilonovae exhibit $\Delta m_z \approx -3.5\text{ to }0$ mag, where larger changes in brightness are associated with longer rise times. Type Ia SNe only exhibit values of $\Delta m_z$ comparable to those of kilonova when the observed rise time is $\sim6$ d (the duration of our search), well beyond the range of observed rise times for kilonovae. When the rise time is $\lesssim6$ d, the average change in brightness is negligible ($\Delta m_z \approx 0.05$ mag). This is because shorter rise times only occur near peak when the light curve evolution is slow. This slow evolution will make Type Ia SNe easy to reject relative to the much more rapidly evolving kilonovae. We note a small fraction of events with $\Delta m_z \approx -2$ mag and rise times of $4-5$ days. These account for $\lesssim0.2\%$ of detected Type Ia SNe and are the result of an event occurring in a range of redshifts where the set of spectral features that produce the second peak of Type Ia SNe appear in {\em z}-band. Thus, Type Ia SNe can be rejected based on $\Delta m_z$, rise time, and $i-z$ color. For the other contaminant populations the separation is not as clean. This is not unexpected since rapid evolution was a strong consideration when constructing our list of potential contaminants. This highlights the critical importance of selecting kilonovae on the basis of their $i-z$ color along with their rapid rise times. 
   
We demonstrated in Figure~\ref{fig:rise} that it is challenging to detect a kilonova during the rise to peak. Therefore, in our simulations we also consider the behavior of sources that are observed in decline. We construct an identical phase-space by replacing the rise time with the decline time, as defined above. Figure~\ref{fig:deccol} shows the decline time-color slice of this phase-space. The layout is identical to Figure~\ref{fig:mcphasecol}. As before, the kilonova population presents $i-z$ colors that are consistently redder than those of the contaminant population with all detected events exhibiting $i-z \gtrsim 0.3$ mag. Furthermore, the fraction of contaminants seen in decline that exhibit $i-z\gtrsim0$ is essentially unchanged compared to those observed on the rise. Additionally, all of the contaminants considered still exhibit $i-z \lesssim 0.3$ mag, so therefore color can still be utilized as a strong discriminant. 

The efficacy of a cut on decline time is diminished compared to a cut on rise time. Both kilonovae and the contaminant population cover the entire distribution of possible decline times from $\sim2-6$ days. A notable exception is Type Ia SNe, which all exhibit a decline time of $\sim6$ days (the duration of the search), because if the source is already in decline at the start of the observations, it will remain detectable for the entire duration of the search. Only $\sim20-25\%$ of kilonovae exhibit a decline time of $\lesssim6$ days, making this an ineffectual cut. On the other hand, the slow evolution is beneficial for separating the Type Ia SNe population from a kilonova if template images at $\gtrsim 10$ days are utilized. In this scenario, the kilonova will have faded away while Type Ia SNe will remain visible allowing them to be easily rejected. 

Figure~\ref{fig:decdm} shows the decline time-$\Delta m_z$ slice of the phase-space. In this context we define $\Delta m_z \equiv m_{z,{\rm last}} - m_{z,{\rm 0}}$, where $m_{z,{\rm last}}$ is the {\em z}-band magnitude from the last observation in which the source is detected. Kilonovae exhibit a change in brightness of $\Delta m_z \approx 0 - 4$ mag, with a weak trend for larger changes in brightness being associated with longer observed decline times. The Type Ia SNe and Pan-STARRS fast transients exhibit an average change in brightness of only $\Delta m_z \approx 0.1$ mag and $\Delta m_z \approx 0.2$ mag, respectively, making the potential for contamination low. However, as with the rise time-$\Delta m_z$ space, the type .Ia and miscellaneous fast transients exhibit changes in brightness that are comparable to those of kilonovae. This makes cuts on $\Delta m_z$ less practical for these sources.  

The key point emerging from our simulations is that $i-z$ color is a strong discriminant for selecting kilonovae, regardless of observing strategy, and whether a kilonova is observed on the rise or decline. In fact, if we consider sources observed during the rise to peak, and we require that $i-z\gtrsim0$ mag we may eliminate $\sim84-87\%$ of detected Type Ia SNe and Pan-STARRS fast transients, $\sim90-94\%$ of detected type .Ia, and $\sim100\%$ of detected miscellaneous fast transients without removing any kilonovae from the selection. If we additionally require that the rise time is $\lesssim4$ days, then we eliminate $\sim94\%$ of detected Type Ia SNe and $\sim98\%$ of detected type .Ia. If we also require that $|\Delta m_z| \gtrsim 0.1$ mag, we eliminate $\sim99\%$ of Type Ia SNe.

Considering sources observed in decline the cuts on $i-z$ color and $\Delta m_z$ remain effective but a cut on decline time does not provide any additional information. Quantitatively, if we apply cuts of $i-z \gtrsim 0$ mag and $|\Delta m_z| \gtrsim 0.1$ mag we eliminate $\sim80\%$ of Type Ia SNe, $\sim60\%$ of type .Ia, $\sim81\%$ of Pan-STARRS fast transients, and $\sim 100\%$ of the miscellaneous fast transients. If we also require that the decline time is $\lesssim 5$ days we eliminate $\sim 100\%$ of the Type Ia SNe and $\sim94\%$ of the Pan-STARRS fast transients, but lose $\sim46\%$ of the kilonova population. However, if additional observations are taken at $\gtrsim10$ days then these kilonovae can be recovered. But, given that cuts made on sources observed to rise are more effective, then in order to maximize the chances of a kilonova detection and identification during a GW follow-up campaign it is essential to obtain both {\em i}- and {\em z}-band data as close to the GW trigger as possible. 

\section{Counterpart Identification In The Absence of a Template} 
\label{sec:diff}
We now investigate the situation where there are no pre-existing template images of the GW error region and we would like to rapidly identify a kilonovae without having to wait for a late-time template. This scenario introduces several challenges. Foremost is the presence of source flux in the image that will be used as a template, which will have an effect on the observables (e.g., brightness, color). Specifically, we can no longer measure the true flux of the source but rather only the difference in flux between the science and template images. In this case, a measurement of the true magnitude of the source is impossible. Without the accurate measurement of source brightness it is also difficult to measure the true color and we therefore lose one of the primary methods of distinguishing kilonovae from the contaminant population. As discussed in detail below, these difficulties can be alleviated by making simple assumptions about the source behavior. 

    \begin{figure*}[t!]
   \centering
	\includegraphics[width=0.7\textwidth]{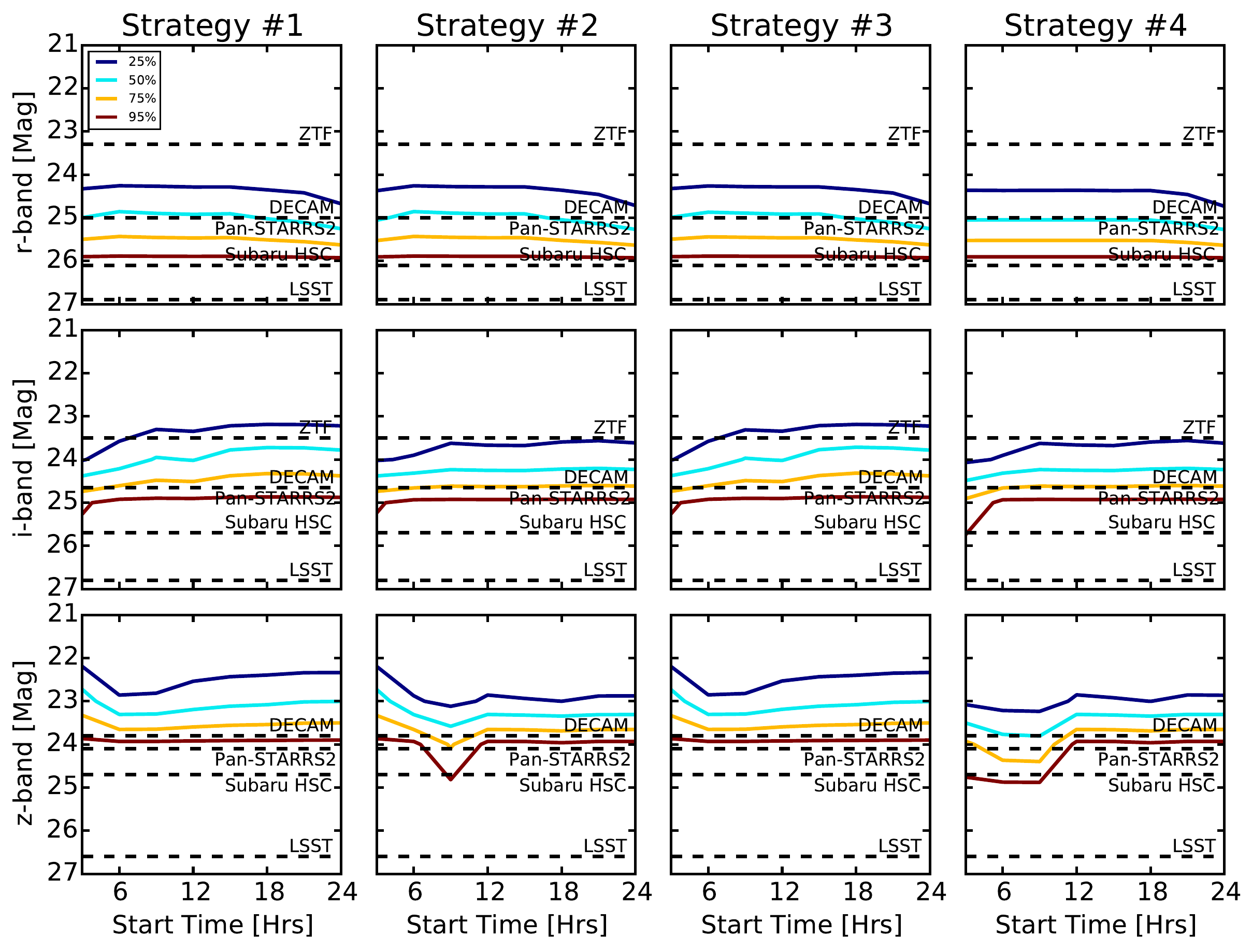}
   \caption{Contours showing the fraction of kilonovae detected as a function of start time and depth for our four observing strategies and three filters when a pre-existing template is not available. We first note that $i$- and $z$-band observations in this scenario must be $\sim1$ mag deeper to achieve detection rates comparable to when a pre-existing template is available. We also note an effect in {\em z}-band where the 95\% contour drops by $\sim1$ mag for observations starting between 6 and 12 hours after trigger using Strategy 2 and for observations starting less than 12 hours after the trigger using Strategy 4. This is the result of the observations straddling the peak of the light curve where the difference in flux between epochs is too small to detect.}
   \label{fig:detdiff}
   \end{figure*}
   
Another obvious challenge is a delay in the timescale of the first science image since the first observation must be used as the template image. This is leads to problems in characterizing  the early behavior of the light curve, as well as in utilizing the rise time as a method of distinguishing kilonvoae from the contaminant population. Using the first observation as a template also reduces the number of available epochs making it more difficult to meet the detection criteria outlined in the previous section. However, these challenges are less severe as they can resolved using a late-time template once potential counterparts are identified.   
   
To address the scenario of using the first image as a template, and to gauge the associated complications, we simulate observations using the same methodology outlined in Section~\ref{sec:MCsims_det}, but taking into account the fact that the first observation is the template. The general procedure is identical to the previous simulations with the following modifications: We now simulate the effects of difference imaging by subtracting the flux in the first observation (the template) from the later observations (the science images) to produce a set of difference fluxes. We account for the noise appropriately by using our search depth as a $5\sigma$ flux limit. We assume that this noise is the dominant source of error and that it is the same for all images. The noise in the difference image is then larger by a factor of $\sqrt{2}$ than that of the individual images. Any source that shows a change in flux with a signal-to-noise ratio (SNR) greater than 5 in the difference image in at least two subtractions is flagged as a detection. If a source exhibits at least one epoch with a positive change in flux, the event is flagged as ``rising."
   
       \begin{figure*}[t!]
   \centering
	\includegraphics[width=0.7\textwidth]{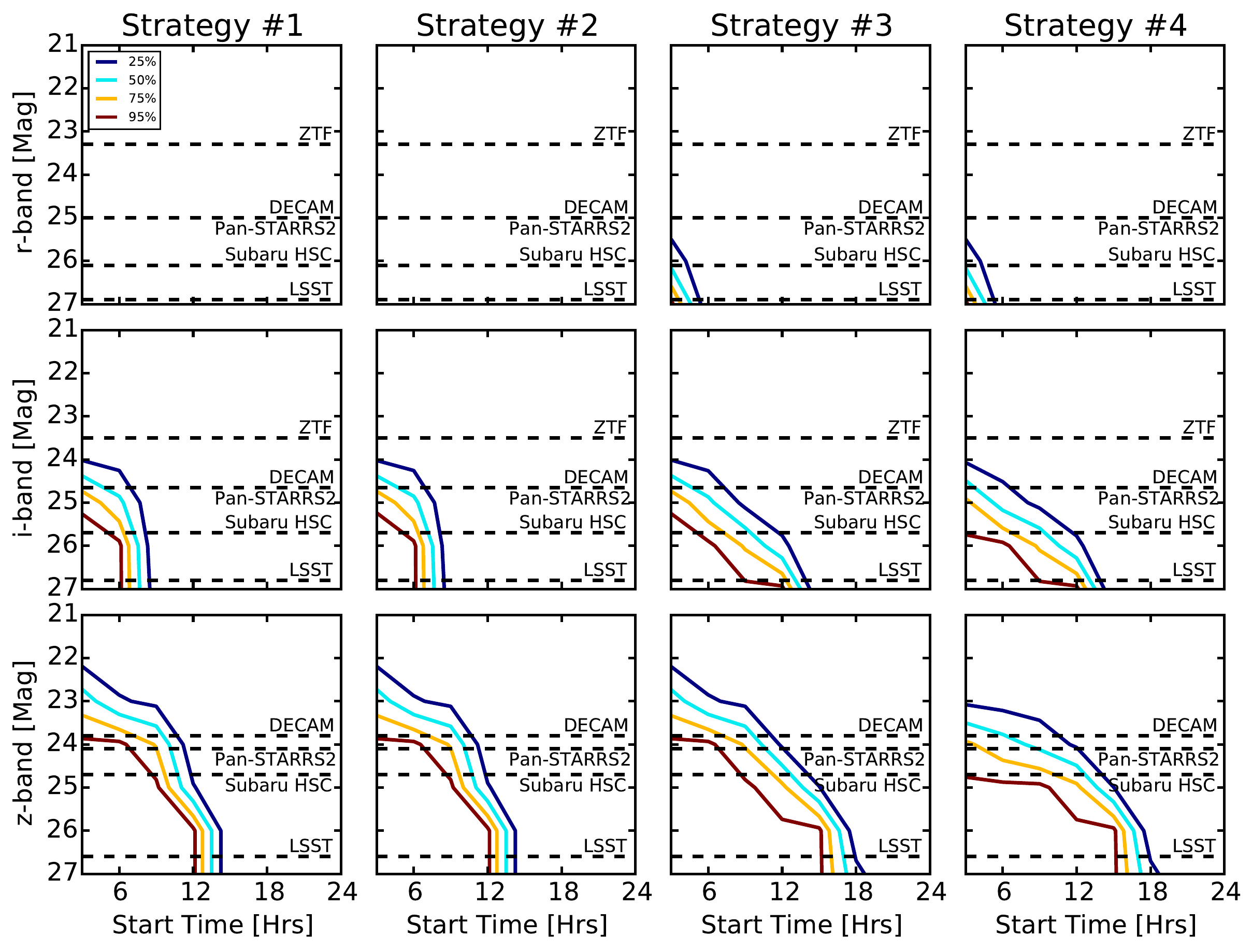}
   \caption{Same as Figure~\ref{fig:detdiff}, but showing the fraction of sources for which brightening is detected. We note that when compared to the contours shown in Figure~\ref{fig:rise} for the case of a pre-existing template, observations must go deeper in this scenario and the detection efficiency quickly tapers off as a function of starting time. This highlights the importance of commencing follow-up observations as soon as possible when a pre-existing template is not available.}
   \label{fig:risediff}
   \end{figure*}

In Figure~\ref{fig:detdiff}, we plot contours of the fraction of sources that pass our detection criterion as a function of start time and search depth. The layout of this plot is identical to that of Figure~\ref{fig:det}. We first note that the {\em r}-band contours for Strategies 1 and 2 are essentially identical to those in Figure~\ref{fig:det}. This is because the $r$-band light curve evolves fast enough between observations that the effects of using the first observation as a template does not hamper detections. However, for Strategies 3 and 4, we no longer find the early-time ($\lesssim6$ hours) boost in detection limit. This is due to the fact that the light curve does not change sufficiently during the three hours, and the source is therefore not detected in the first difference image. There is therefore no benefit from the additional observation on the first night. On the other hand, in {\em i}-band we find that on average the depth required to reach a certain detection rate is $\sim 1$ mag deeper than in the case of a pre-existing or late-time template (Figure~\ref{fig:det}). Specifically, for Strategy 1 the 50\% contour starts at $\approx24.4$ mag for $t_{{\rm start}} \sim 3$ hours and rises to $\approx 23.7$ mag by $t_{{\rm start}} \sim24$ hours. In Strategy 2, the 50\% contour is flat at $\approx24.3$ mag. The 95\% contour is flat at $\approx25$ mag in both cases. Strategy 3 is identical to Strategy 1, again highlighting the fact that the rapid cadence on the first night does not boost the detection rate. Lastly, we find that the rapid early-time cadence combined with the slow late-time cadence of Strategy 4 actually produces a drop in efficiency for observations beginning at $\sim$ 3 hours where the 95\% contour dips to $\approx 25.7$ mag.

       \begin{figure*}[t!]
   \centering
	\includegraphics[width=0.7\textwidth]{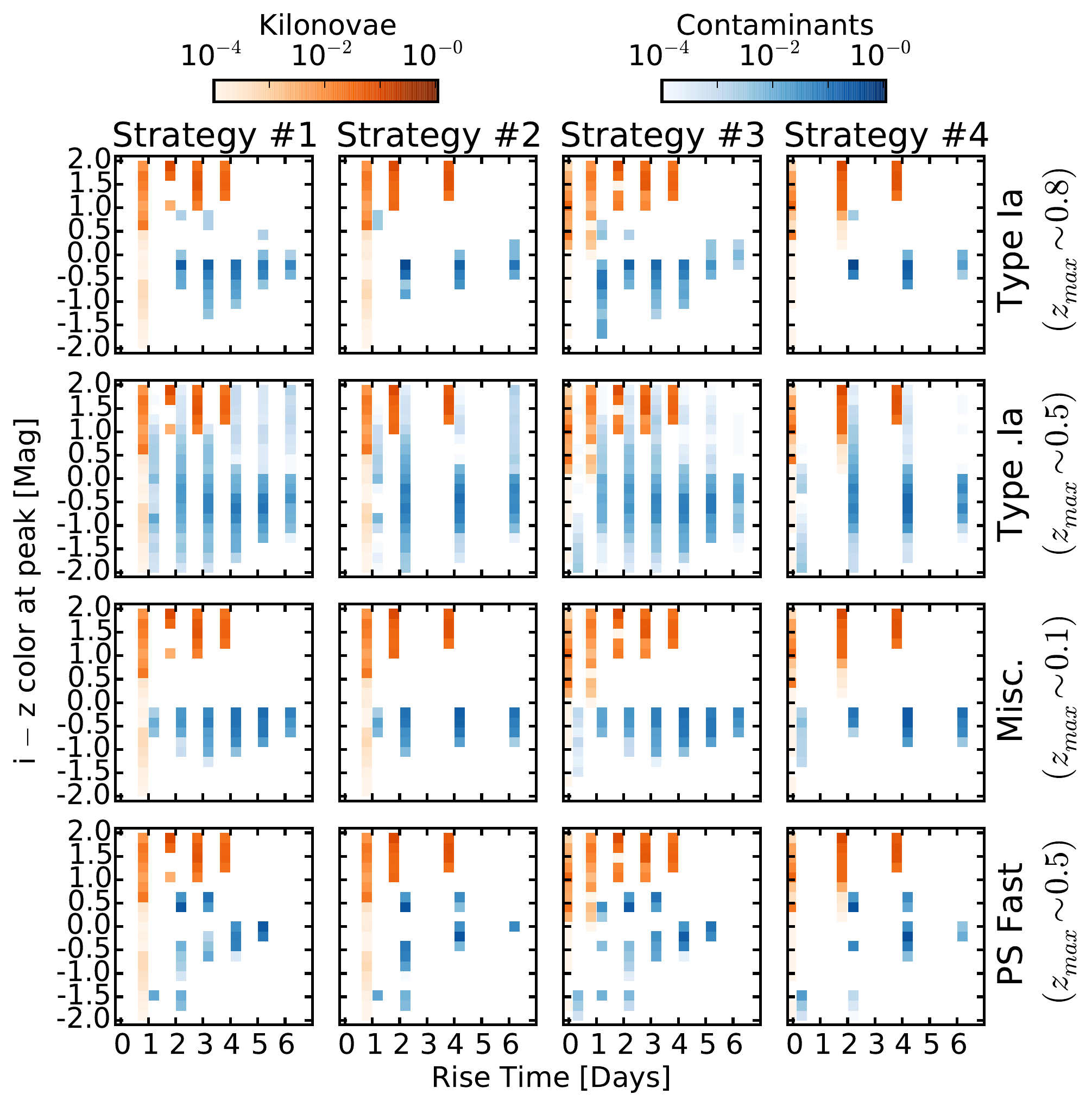}
   \caption{The rise time-color slice of our phase-space when a pre-existing template is not available. We note that if the rise time is long ($\gtrsim2$ days), then there is sufficient evolution in the light curve that colors can be measured with reasonable accuracy and kilonovae can be separated from the contaminants as before. However, when the rise time is short $(\sim 1$ day), then the change in flux is small and the colors are not well constrained.}
   \label{fig:phaserisediff}
   \end{figure*}

The most pronounced change relative to the case of a pre-existing template (Figure~\ref{fig:det}) can be seen in the {\em z}-band contours. In Strategy 1, the 50\% (95\%) contour is approximately $\approx 23\,(23.9)$ mag, which is again $\sim 1$ mag deeper than for the case of a pre-existing template. Strategy 2 shows a change of behavior in the 95\% contour between 6 and 12 hours, where the required depth dips to $\approx 24.8$ mag. This is because observations starting during this time frame will be roughly symmetrical about the peak of the light curve. Therefore, the change in flux is small and the source will not be detected in the difference image. The slow cadence of Strategy 2 then makes it difficult to obtain enough observations to reach our threshold for a detection. This effect is also seen in Strategy 4 where the 95\% contour dips to $\approx 24.8$ mag for any starting time less than 12 hours. These contours are summarized in Table~\ref{tab:det}.

In Figure~\ref{fig:risediff}, we show contours for the fraction of sources with a detected brightening between the template and at least one of the science images. These sources must also pass the detection criterion described above. The layout is identical to that of Figure~\ref{fig:rise}. The behavior of the contours is different compared to the case of a pre-existing template for all bandpasses and strategies. In Figure~\ref{fig:rise}, the contours were flat and then exhibited a sharp drop off when the first observation began post peak. Here, the contours show a steady decline as a function of start time. This is because as the start time of the first observation (i.e., the template) approaches the peak, the observed change in flux will decrease. Therefore, it becomes progressively more difficult to detect a brightening and the required depth increases.

Specifically, for {\em r}-band, no brightening is detected when utilizing Strategies 1 and 2. The same is effectively true for Strategies 3 and 4, unless the observations reach a depth of $\approx 26.2$ mag and begin $\lesssim$ 3 hr after the GW trigger. Thus, even with deep Subaru HSC or LSST observations, brightening could only be detected in $\sim25-50$\% of sources for observations starting within 3 hours of the GW trigger. In {\em i}-band, for Strategies 1 and 2, the 50\% (95\%) contour begins at $\approx 24.4$ (25.3) mag and no rise is detected if $t_{{\rm start}} \gtrsim 7$ (6) hours. For Strategies 3 and 4, the 50\% (95\%) contours start at $\approx 24.5\,(25.3)$ mag and fall below the LSST limiting magnitude by $\sim 9\,(6)$ hours. Thus, the detection of a rise in $i$-band is not feasible with existing telescopes without a pre-existing template. Lastly, for {\em z}-band, the 50\% (95\%) contours in Strategies 1 and 2 begin at $\approx 22.7\,(23.9)$ mag and fall below the LSST limiting magnitude by $\sim 15\,(12)$ hours. For Strategy 3, the required depths are unchanged compared to Strategies 1 and 2, but the 50\%(95\%) contours extend out to $\sim 17\,(15)$ hours. For Strategy 4, the required depth to achieve a 50\% (95\%) detection rate is $\sim 1$ magnitude deeper compared to Strategy 3, but the required start times are identical. As above, this is possible with Subaru HSC and LSST, but other instruments (e.g., DECam, Pan-STARRS2) will only be able to achieve a 95\% detection rate for $z$-band observations starting within $\sim6$ hours of the GW trigger. This highlights the increased difficulty in characterizing the early-time behavior of the light curve because we lose information from the first epoch when it is used as the template image. We note that this information can be recovered after the fact using a late-time template but here we are concerned with rapid identification.

\subsection{Contaminant Rejection In The Absence of a Template}
We now investigate how the absence of a pre-existing template affects our ability to reject contaminants. We carry out our simulations in the following way. The flux in the first observation (the template) is subtracted off from the later observations. Noise calculations are handled identically to those outlined above for the detectability simulation, and we impose the same detection criterion for events. The rise/decline time phase-space coordinates are computed using the same definition as in Section~\ref{sec:MCsims_cont}. The challenge then is computing an analogue for the values of the $i-z$ color and $\Delta m_z$ when the source flux is not known absolutely (due to the presence of source flux in the template image). 

When there is source flux in the template image we are only able to measure a difference in flux between the template and science image. We denote this as $\Delta f_{\phi} \equiv f_{s,\phi} - f_{t,\phi}$ where $f_{s,\phi}$ is the flux observed in the science image and $f_{t,\phi}$ is the flux observed in the template image, both in some bandpass $\phi$. We can then define an AB magnitude using the standard formula $m^{\star}_{\phi} = -2.5\log{|\Delta f_{\phi}|}-48.6$, where the star denotes magnitudes computed from a difference in flux. The $\phi_1 - \phi_2$ color for bandpasses $\phi_1$ and $\phi_2$ is then $m^{\star}_{\phi_1} - m^{\star}_{\phi_2}$. In the case that $|\Delta f_{\phi}|$ is large, and consequently the detection is of high statistical significance,  we approach the limit where the source flux in the template is negligible and a reasonable approximation for the true magnitude and color of the source can be obtained. However, for less significant detections, the measured color will not be an accurate measurement of its true value.

Lastly, we construct a straight-forward analogue to the rise/decline time-$\Delta m$ slice of our phase-space. In the case of a rising source, we compute $m^{\star}_z$ using the largest positive difference in flux between the template image and the science images. If the source is observed only in decline then the largest negative difference in flux is used. This will allow us to understand how rapidly a source evolves relative to others in this study (e.g., kilonovae versus contaminants).

    \begin{figure*}[t!]
   \centering
	\includegraphics[width=0.7\textwidth]{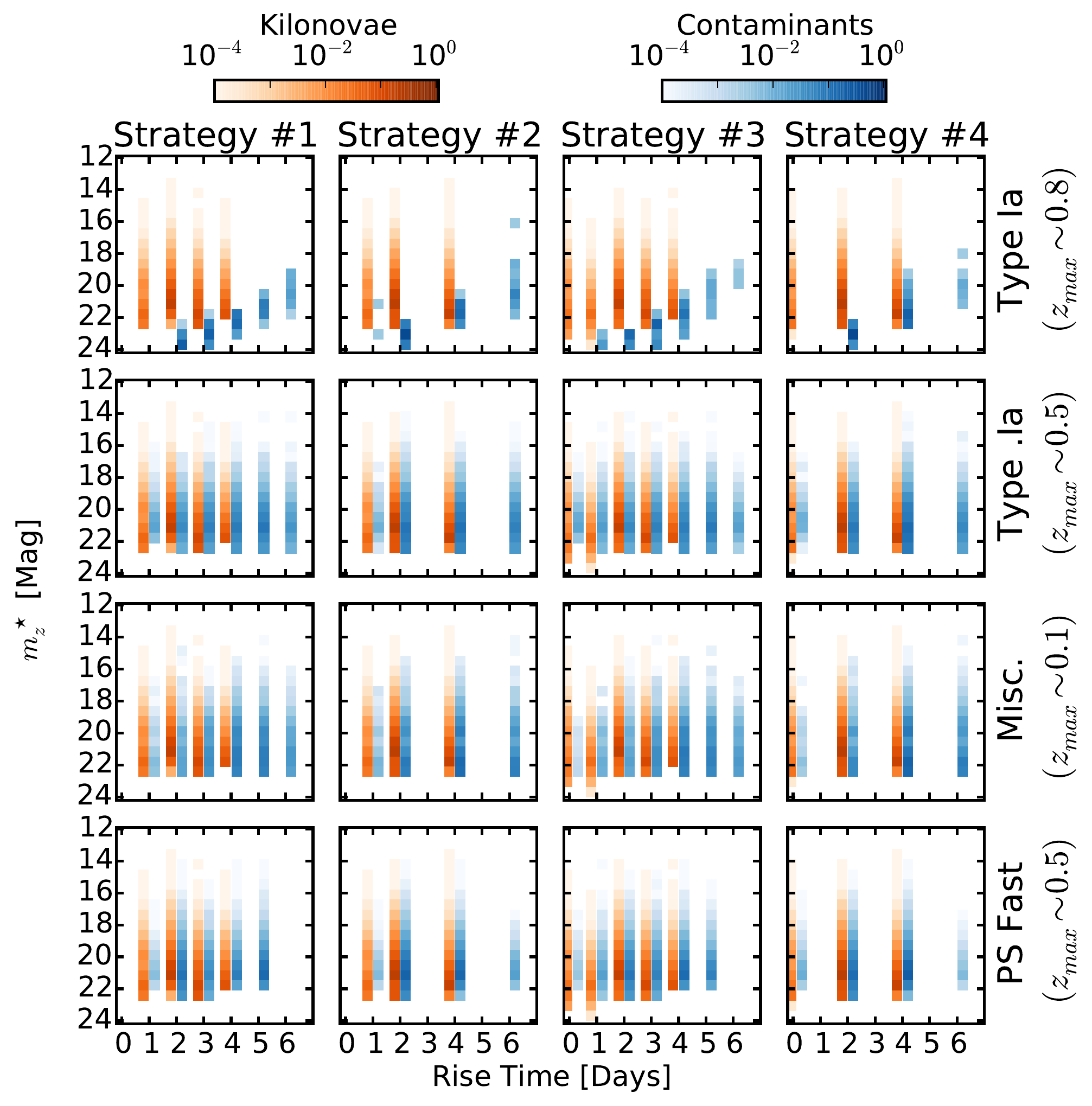}
   \caption{The rise time-$m^{\star}_z$ slice of our phase-space. The slower evolution of Type Ia SNe is apparent. For the remaining contaminants the observed change in flux is comparable to that observed for kilonovae. The rise times exhibited by both the kilonovae and contaminant populations are also comparable, however only contaminants show a rise time $\gtrsim 5$ days.}
   \label{fig:phaserisediff_df}
   \end{figure*}
   
In Figure~\ref{fig:phaserisediff} we plot the rise time-color slice using the same layout as in Figure~\ref{fig:mcphasecol}. Inspecting the kilonovae across all four observing strategies we note that when the observed rise time is small ($\lesssim$ 1 day, i.e. when the difference in time between the template and first science image is small) then the color is not well constrained due to $|\Delta f|$ being too small to obtain an accurate measurement. However, the fraction of kilonovae with a rise time of $\lesssim1$ day is only $\sim5-8\%$. We also note that the fraction of detected kilonovae with $t_{{\rm rise}} \lesssim 1$ day that exhibit $i-z\approx -2$ to $+0.5$ mag (i.e., in the range that overlaps with contaminants) is at most $\sim1\%$ for Strategies 3 and 4, and $\lesssim1\%$ for Strategies 1 and 2. More importantly, when the observed rise time is longer $(\gtrsim 2$ days), then $|\Delta f|$ is large enough that the $i-z$ color can be well constrained and it is generally $i-z\gtrsim1$ mag. This is consistent with the color measured when a pre-existing template is available (Figure~\ref{fig:mcphasecol}). The fraction of detected kilonovae that present a rise time of $\gtrsim 2$ days is $\sim45-55\%$. This fraction is dominated by kilonovae models with an ejecta mass of $\Mej = 10^{-1} \;M_{\odot}$ and $\beta_{\text{ej}} = $ 0.1 or 0.2.

Turning to the contaminant population, we find that several of the source classes have poorly constrained colors with some events exhibiting $i-z > 0$ mag. We can assess the impact of these contaminants by utilizing the rate data in Tables~\ref{tab:rates} and~\ref{tab:rates_Iae}. This is achieved by multiplying the expected fraction of sources in a region of the phase-space with the number of sources expected to occur during the search. This gives the actual number of sources detected. As expected, Type Ia SNe are the most numerous contaminant with $\sim3-12$ sources expected to exhibit $i-z\gtrsim0$ mag. Strategies 3 and 4 produce the fewest number of Type Ia SNe detections ($\sim 3-5$), while Strategy 2 produces the most ($\sim 12$). For the Pan-STARRS fast transients, we find that the number of sources exhibiting $i-z\gtrsim0$ mag is $\sim3$ across all four strategies. Lastly, we expect $\ll1$ type .Ia and miscellaneous fast transients to exhibit $i-z\gtrsim0$ mag. If we additionally require that the rise time is $\lesssim 4$ days the number of detected Type Ia SNe is reduced to $\sim1-7$. This does not affect the number of detected Pan-STARRS fast transients. Ultimately, we do not expect the lack of a pre-existing template to significantly hinder our ability to separate kilonovae from contaminants on the basis of their $i-z$ color and timescale, if the source is observed during the rise to peak. 

Figure~\ref{fig:phaserisediff_df} shows the rise time-$m^{\star}_z$ slice of our phase-space. We first note the wide range of values for $m^{\star}_z$ that appear in our simulations. This arises from considering contaminants over a wide redshift range or in the case of the kilonovae, over a range of model parameters. We further note that there does not appear to be an obvious cut utilizing $m^{\star}_z$ that helps to distinguish kilonovae from the contaminant population. However, we can separate kilonovae from the contaminant populations by utilizing a cut on timescale as all of the kilonovae exhibit a rise time of $\lesssim 4$ days. Therefore, if we require that sources exhibit a rise time of $\lesssim 4$ days, then we expect to eliminate $\sim 100$ Type Ia SNe, $\sim 1$ Pan-STARRS fast transient, and $\ll1$ type Ia and miscellaneous fast transients, without rejecting any kilonovae.

    \begin{figure*}[t!]
   \centering
	\includegraphics[width=0.7\textwidth]{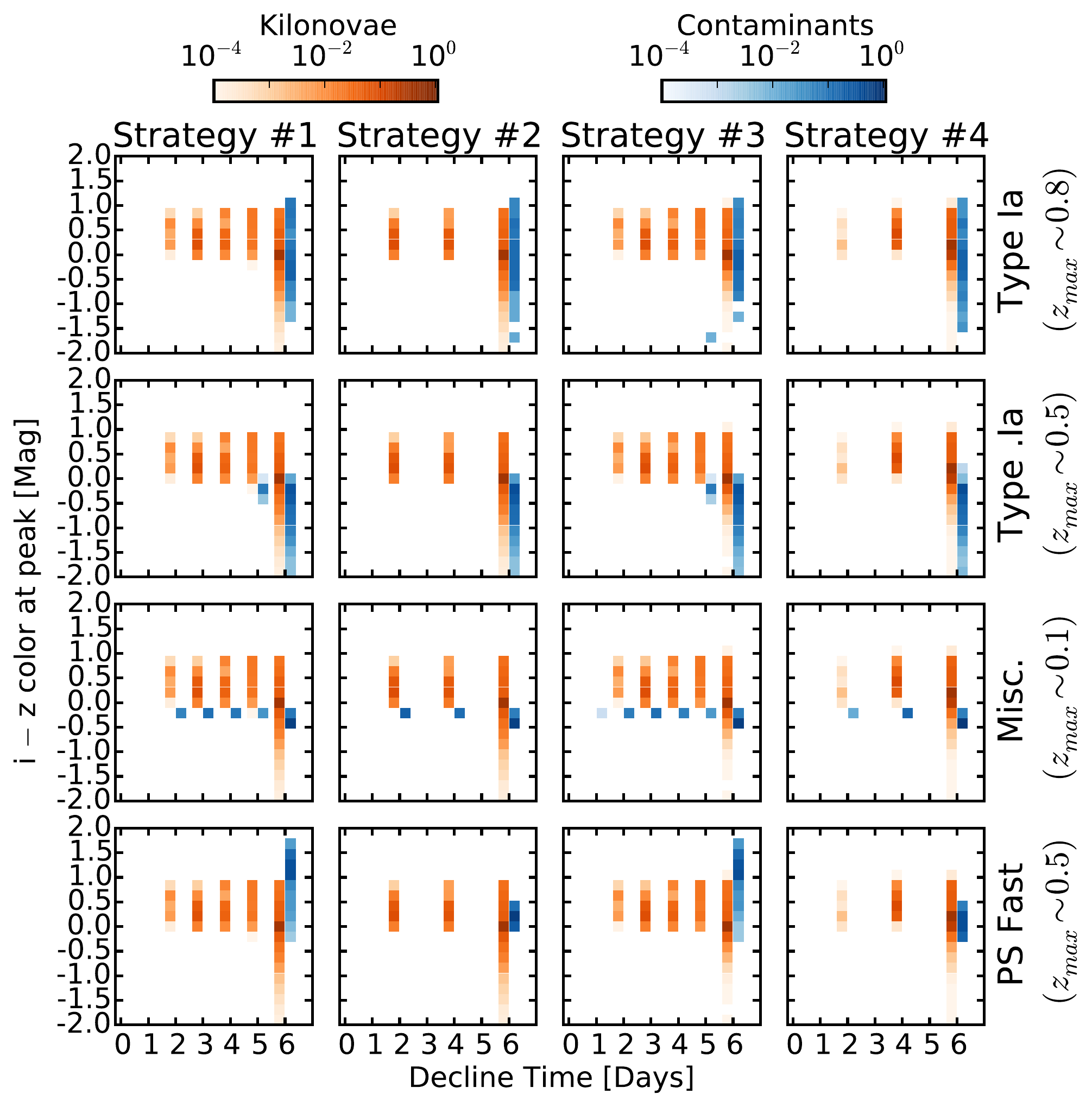}
   \caption{Same as Figure~\ref{fig:phaserisediff} but showing the decline time-color slice of the phase-space. The kilonovae are not always clearly separated from the contaminant population due to the imprecise color measurements. However, contaminants (with the exception of the miscellaneous fast transients) do stand out from kilonovae on the basis of their long decline times ($\gtrsim 6$ days, the maximal length given our search duration).}
   \label{fig:phasedecdiff}
   \end{figure*}
   
Despite the lack of a clear separation between the kilonovae and contaminant populations in $m^{\star}_z$, the requirement that sources exhibit a $m^{\star}_z$ greater than $5\sigma$ in the difference images does reduce the overall number of contaminants detected. Utilizing this criterion, the detection rates for Type Ia SNe and Pan-STARRS fast transients (the two most numerous contaminants) on the rise is $\sim1\%$ for both populations, down from $\sim60\%$ and $\sim25\%$, respectively, in the case of a pre-existing template. The downside of this approach is that $\sim10-15\%$ of kilonovae are also lost. These sources can be recovered with a late-time template, but they will not be detected in real time. 

In Figure~\ref{fig:phasedecdiff} we plot the decline time-color slice of our phase-space. We find that there is no longer a clear separation in color between the kilonovae and contaminant populations. This is true for all contaminants and across all observing strategies. This effect is the result of the {\em apparent} slow evolution of the light curve during the decline phase relative to the template flux. This can occur if the observations straddle the light curve peak or if the source simply evolves slower after peak. However, we can separate kilonovae from the contaminant populations by utilizing a cut on timescale as all of the contaminants predominantly exhibit a decline time of $\gtrsim 6$ days (the maximal value given the duration of our search). Therefore, if we require that sources exhibit a decline time of $\lesssim 5$ days, then we expect no Type Ia SNe, type .Ia and Pan-STARRS fast transients, and $\ll 1$ miscellaneous fast transient to be detected, but this cut will reject $\sim15-20\%$ of kilonovae.

    \begin{figure*}[t!]
   \centering
	\includegraphics[width=0.7\textwidth]{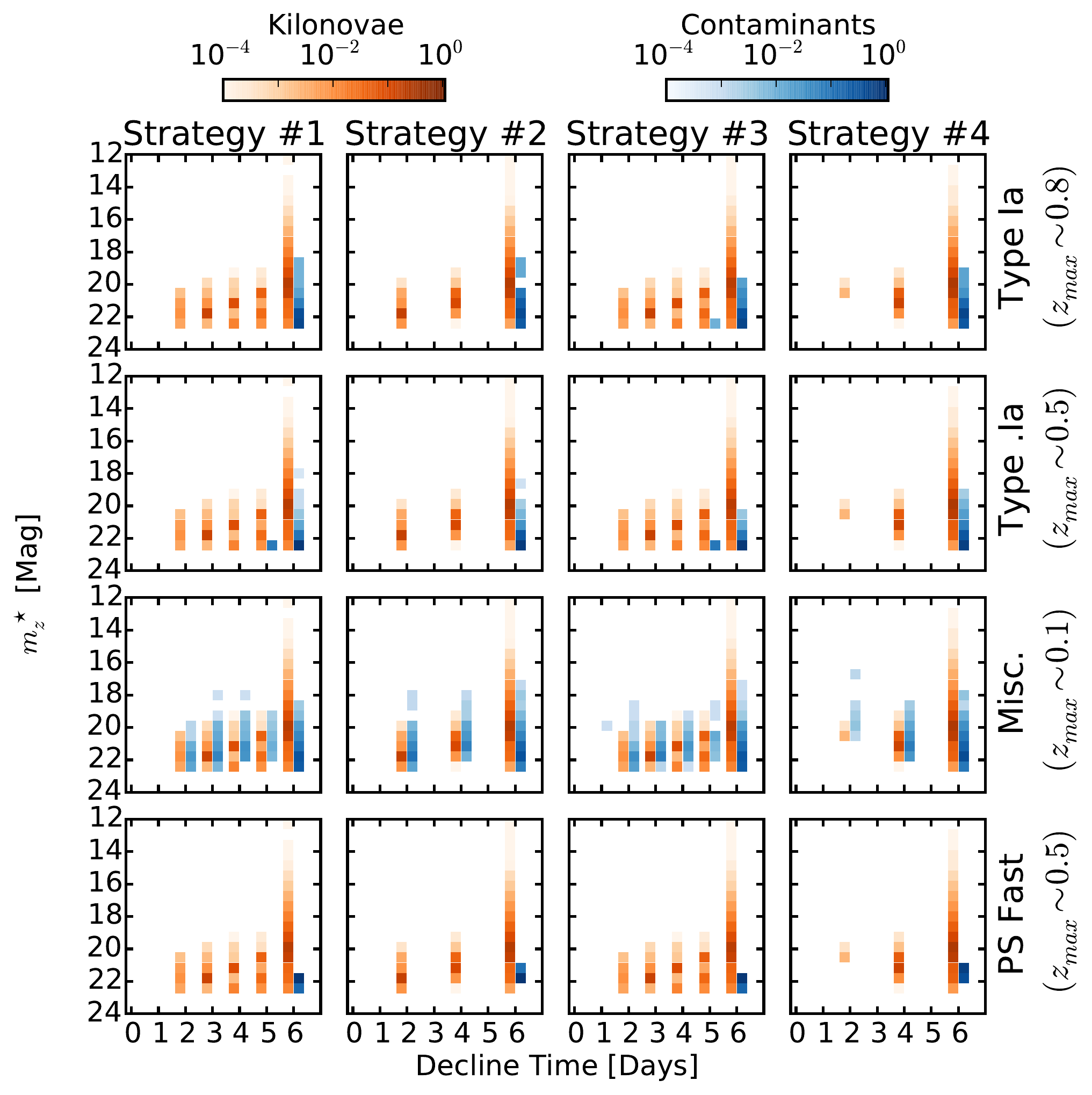}
   \caption{Same as Figure~\ref{fig:phaserisediff_df} but showing the decline time-$m^{\star}_z$ slice of the phase-space. As in Figure~\ref{fig:phaserisediff_df}, the contaminants are not clearly separated from the kilonovae population in $m^{\star}_z$. However, as in Figure~\ref{fig:phasedecdiff}, the long decline times can be used to distinguish kilonovae from the contaminant populations.}
   \label{fig:phasedecdiff_df}
   \end{figure*}
   
Figure~\ref{fig:phasedecdiff_df} shows the decline time-$m^{\star}_z$ slice of the phase-space. As in the case of sources detected on the rise, there is no clear cut on $m^{\star}_z$ that will help to separate the kilonovae from any contaminants. Furthermore, a cut on decline time (e.g., $\gtrsim 6$ days) does not yield any new information beyond what is gained from the decline time-color slice of the phase-space. However, as before, utilizing a detection criterion on $m^{\star}_z$ does reduce the total number of contaminants detected. This effect is enhanced for sources such as Type Ia SNe and the Pan-STARRS fast transients, as they show slower post-peak evolution in their light curves. Here, the detection rates for these sources, when seen on the decline, is $\ll1\%$ in both cases. This is down from $\sim20\%$ and $\sim15\%$, respectively, in the case of a pre-existing template. However, the detection criterion on $m^{\star}_z$ reduces the kilonovae detection rates by $\sim5\%$. 

We now summarize the efficacy of cuts based on $i-z$ color, timescale, and change in flux for separating kilonovae from the contaminants when a pre-existing template is not available. For sources observed during the rise to peak, requiring $i-z\gtrsim0$ mag eliminates $\sim98\%$ of detected Type Ia SNe, $\sim91\%$ of detected type .Ia, $\sim100\%$ of detected miscellaneous fast transients, and $\sim 56\%$ of detected Pan-STARRS fast transients while eliminating $\lesssim0.1\%$ of detected kilonovae. If we additionally require that the rise time is $\lesssim4$ days, then we eliminate $\sim99\%$ of detected Type Ia SNe, $\sim94\%$ of detected .Ia SNe, and $\sim59\%$ of detected fast Pan-STARRS transients, without altering the fraction of kilonovae eliminated. In this scenario, an additional cut on $m^{\star}_z$ does not yield any additional information.

For sources observed only in decline, a cut of $i-z\gtrsim0$ mag eliminates $\sim50\%$ of detected Type Ia SNe, $\sim99\%$ of detected type .Ia, and $\sim 100\%$ of detected miscellaneous fast transients. The fraction of detected Pan-STARRS fast transients eliminated depends on the observing strategy; for Strategies 1 and 3 it is essentially zero, while for Strategies 2 and 4 it is $\sim11\%$ and $\sim37\%$, respectively. The reason for this variation is the fact that the different observing cadences are better matched to specific subsets of the population. The cut of $i-z\gtrsim0$ mag color also eliminates $\sim 15\%$ of detected kilonovae. If we additionally require that the decline time is $\lesssim 5$ days then we remove $\sim100\%$ of the contaminant population but also eliminate $\sim60-80\%$ of detected kilonovae. No additional information is gained from a cut on $m^{\star}_z$. Therefore, cuts on $i-z$ color and timescale are less effective when the source is observed on the decline only. This highlights the need for the rapid triggering of follow-up observations.
  
Despite these challenges, the best approach in the absence of a pre-existing template is unchanged from the results of Section~\ref{sec:MCsims_cont}. The optimal follow-up cadence is still Strategy 1, which allows good sampling of the light curve while maximizing the area that can be covered in a single night of observations. This strategy is also not subject to the dip in detection efficiency as a function of start time, seen in Strategies 2 and 4. However, we note that the previously recommended depths of 24 AB mag in {\em i}-band and 23 AB mag in {\em z}-band are only expected to detect $\sim50-75\%$ of kilonovae in this scenario (Figure~\ref{fig:detdiff}). The $i-z$ color still provides a strong constraint for separating kilonovae from contaminants but it is most reliable when the kilonova can be observed during the rise to peak. If this is not feasible due to other observational constraints, then it is best to resort to obtaining template images once the kilonova has faded in the optical ($\gtrsim$ 10 days). 

\section{Effects of Alternative Kilonovae Models}
\label{sec:altkilo}

So far, we have only considered kilonova models affected by the large opacities of r-process heavy nuclei. We now consider the effects of modifications to the standard kilonova models and their impact on our detectability results. The key issue in determining the nature of kilonova emission is the electron fraction, $Y_e$. The value of $Y_e$ in the dynamical ejecta and from accretion disk winds is low $(Y_e \lesssim 0.2)$ which leads to the production of r-process nuclei and strong suppression of the optical emission. However, if the electron fraction of the ejecta is above the critical value of $Y_e \approx 0.25$, r-process nucleosynthesis is not able to produce heavy elements  (particularly the lanthanides, Kasen et al. 2014). The resulting opacities will be closer to those from the Fe-peak elements, leading to a bluer transient. Here we investigate speculative scenarios that could produce a ``blue" kilonova component and assess their impact on {\em r}-band detectability.

\subsection{Neutrino Driven Winds}
\label{sec:neutrino}
Ejecta can be produced by the outflows from a post-merger accretion disk (Dessart et al. 2009, Fern\'{a}ndez \& Metzger 2013, Fern\'{a}ndez et al. 2014). Fern\'{a}ndez \& Metzger (2013) performed numerical two-dimensional hydrodynamical simulations of accretion disks and found that $\sim$10\% of the initial disk mass is ejected over timescales of $\sim1$ s. These ejecta are a neutron-rich wind with an electron fraction of $Y_{\text{e}} \sim 0.2$, producing conditions that are favorable for r-process nucleosynthesis. This type of outflow will therefore not drastically alter the observational properties of the kilonova. 

Recent work by Fern\'{a}ndez et al. (2014) found that the amount of ejected mass is a function of the spin of the remnant black hole, with more rapidly spinning black holes leading to higher ejecta masses by a factor of several. Furthermore, they found that $Y_{{\rm e}}$ increased monotonically with black hole spin due to an increased early-time neutrino luminosity from a rapidly spinning black hole. They found that in the most extreme cases $(a \gtrsim 0.8)$, the system could eject $\sim10^{-5} - 3\times10^{-3}\;M_{\odot}$ of material with $Y_{{\rm e}} \gtrsim 0.25$. Specifically, an ejecta mass of $\sim10^{-3}\;M_{\odot}$ will produce a blue component in the kilonovae light curve with $\nu L_{\nu} \approx 7.5\times10^{40}$ erg s$^{-1}$ at $\sim6500$ \AA ($r \approx 22.5$ mag at 200 Mpc) and a peak time of $\lesssim$ 1 day (Kasen et al. 2014). This is significantly brighter in $r$-band than the kilonovae models discussed previously, but the short timescales make observing this component challenging.

\subsection{A Long-Lived Hyper-massive Neutron Star}
\label{sec:HMNS}
Another modification can be caused by a  surviving hyper-massive neutron star (HMNS). The HMNS is a potential result of NS-NS mergers, but in simulations it often collapses rapidly $(t \lesssim 100$ ms) to form a black hole (Sekiguchi et al. 2011). The prompt formation of a black hole is an implicit assumption in the kilonova models discussed so far in this paper. However, if the total mass of the binary is less than a critical value, the HMNS can survive for a longer period of time (Bauswein et al. 2013a, Kaplan et al. 2013). In this scenario, the resulting neutrino luminosity will raise the electron fraction of the disk wind ejecta and will suppress r-process nucleosynthesis. Thus, the lanthanide-free disk wind material will produce a brighter, bluer, and shorter lived component in the kilonova light curve (e.g., Metzger \& Piro 2014, Metzger \& Fern\'{a}ndez 2014, and references therein). This emissions component will have a strong viewing angle dependence, with the polar regions being more strongly irradiated by neutrinos than the equatorial regions which will remain obscured by lanthanide-rich material. Consequently, the blue early component is more pronounced when the merger is observed along the polar axis. 

Metzger  \& Fern\'{a}ndez (2014) found that the luminosity and timescale of this initial blue component are strongly dependent on the lifetime of the HMNS, with typical values ($\sim 100$ ms) leading to peak luminosity that is a factor of a few times brighter than the red kilonova component, with a timescale of a few days for emission along the polar direction. The light curves of this early-time component were computed in detail using wavelength-dependent radiative transfer calculations by Kasen et al. (2014). They found that for an HMNS with a lifetime of $\sim100$ ms, the initial blue component will have a peak luminosity of $\nu L_{\nu} \approx 1.3\times10^{41}$ ergs s$^{-1}$ at $\sim6500$ \AA\,  ($r \approx 22$ mag at 200 Mpc) and a peak time of $\sim2$ days. Kasen et al. (2014) also compared the blue component to the r-process powered emission from a $10^{-2}\;M_{\odot}$ torus of dynamical ejecta. They found that the blue component will be visible for polar viewing angles $(\theta \lesssim 15$ deg). The fraction of sources expected to be observed at this viewing angle is $\sim7\theta^2\sim0.48$ (Metzger \& Berger 2012). However, at these viewing angles the kilonova emission will be overwhelmed by the optical afterglow of an on-axis SGRB that may accompany the merger. Furthermore, if the source is observed at large viewing angles $(\theta \gtrsim 15$ deg) then the emission will become obscured by the dynamical torus and the $r$-band emission will then peak at $\gtrsim 23.7$ mag ($\sim2$ mag fainter than the polar case for a torus mass of $10^{-2}\;M_{\odot}$).

       \begin{figure}[h!]
   \centering
	\includegraphics[width=\columnwidth]{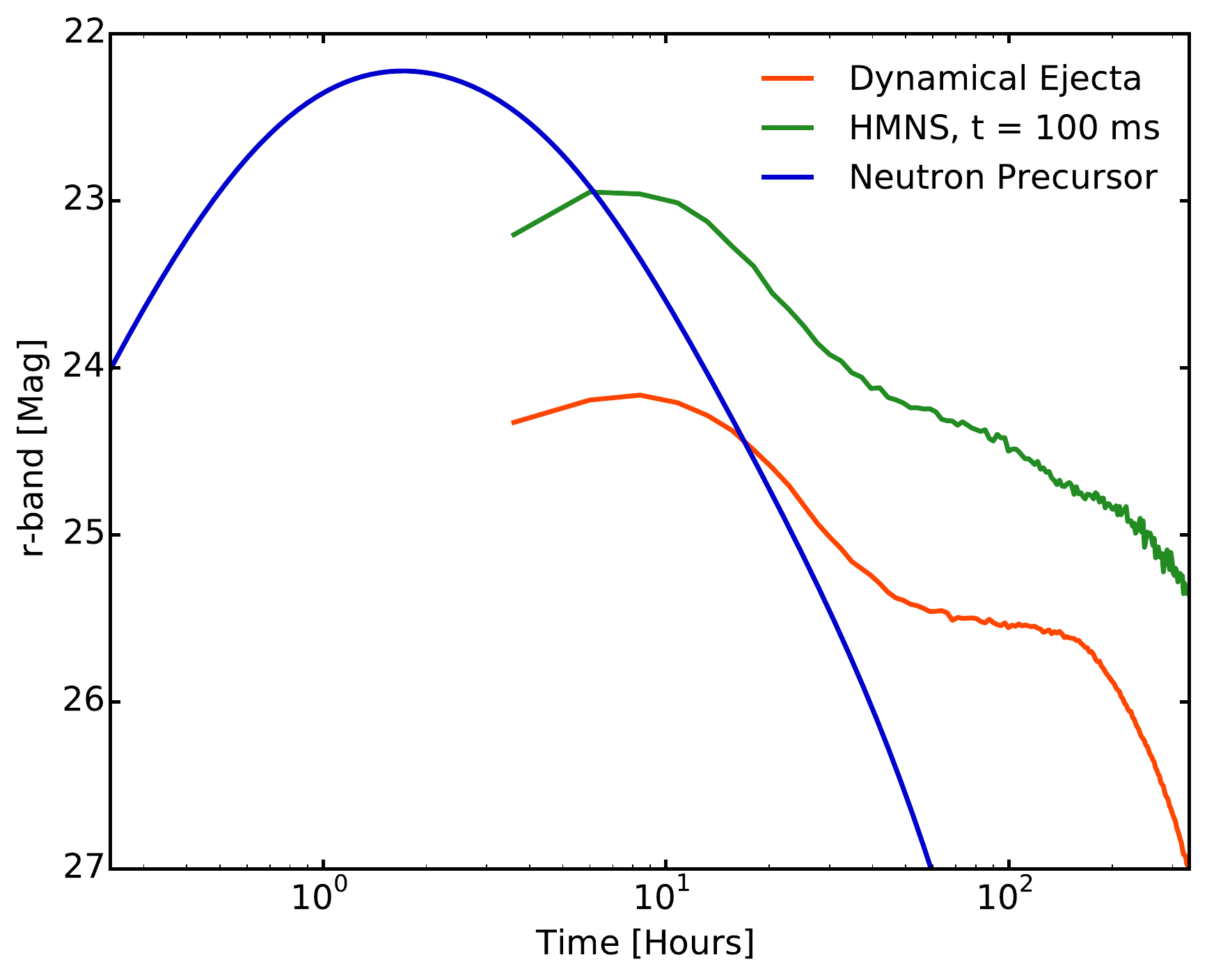}
   \caption{{\em r}-band light curves from the alternative kilonova models discussed in Sections~\ref{sec:HMNS} and ~\ref{sec:neutronpre}. The {\em r}-band light curve for the neutron-rich dynamical ejecta with $M_{{\rm ej}} = 10^{-2}\;M_{\odot}$ and $\beta_{{\rm ej}} = 0.2$ is plotted for comparison. The alternative models produce light curves which can be significantly brighter and bluer than the dynamical ejecta but peak at earlier times.}
   \label{fig:altLC}
   \end{figure}
   
          \begin{figure*}[t!]
   \centering
	\includegraphics[width=0.7\textwidth]{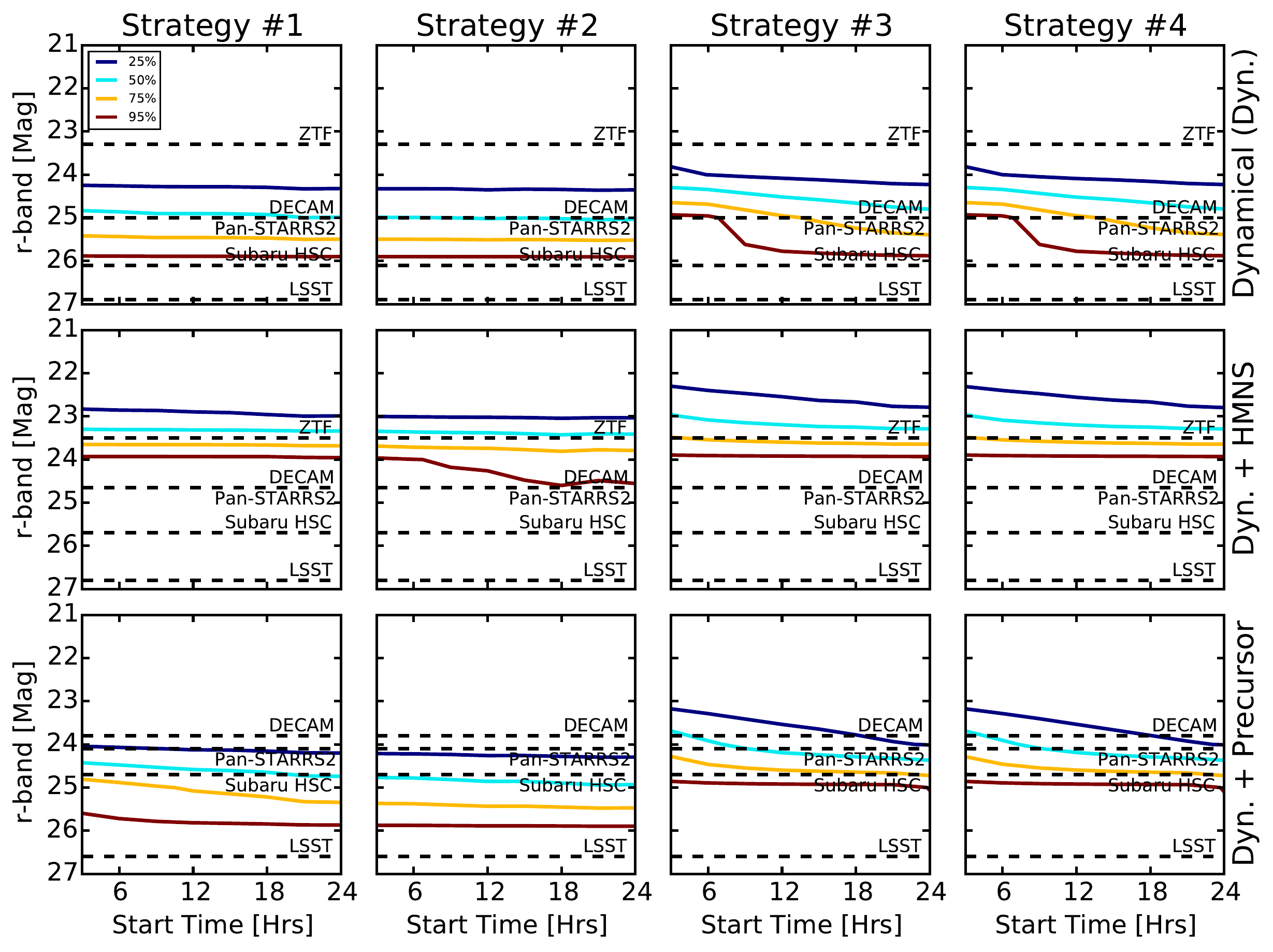}
   \caption{Contour plots for the {\em r}-band detectability of alternative kilonovae models. The top row shows the contours for the dynamical ejecta (reproduced from Figure~\ref{sec:MCsims_det}). The middle row shows contours for a long-lived HMNS with a lifetime of $\sim 100$ ms, surrounded by a dynamical tours with a mass of $10^{-2}\;M_{\odot}$  of neutron-rich material observed at a viewing angle of $\theta \sim 18$ deg as modeled by Kasen et al. (2014). The bottom row shows the effect of an early-time neutron precursor as modeled by Metzger et al. (2014).}
   \label{fig:altdet}
   \end{figure*}
   
\subsection{A Neutron Precursor}
\label{sec:neutronpre}
Lastly, we consider the case in which a small fraction of the neutron-rich ejecta is expanding so rapidly that it avoids the initial phase of {\em r}-process nucleosynthesis. This effect has been seen in several general relativistic simulations (Bauswein et al. 2013b, Goriely et al. 2014, Just et al. 2014). In particular, Bauswein et al. (2013b) found that for a small fraction of the ejecta $(\sim 10^{-5} - 10^{-4}\; M_{\odot})$, the expansion timescale was sufficiently rapid $(\lesssim 5$ ms) that they evade neutron capture. These neutrons will undergo $\beta$-decay and release energy into the ejecta. Given that the photon diffusion time is on the order of a few hours in the outer layers of the ejecta, a significant fraction of this energy will manage to produce radiation. The light curves of this ``neutron precursor" were modeled by Metzger et al. (2014). They found that the $\beta$-decay from the population of free neutrons at the leading edge of $10^{-4}\;M_{\odot}$ of ejecta can produce a transient with a timescale of $\sim1-2$ hours and a peak {\em r}-band magnitude of $\approx 22.2$ mag at 200 Mpc.

\subsection{Implications for {\em r}-band detectability}
We now investigate the effect of these speculative models on the early-time detectability of kilonovae using {\em r}-band observations. We construct the light curves for the long-lived HMNS from the time-resolved spectra modeled by Kasen et al. (2014) using the procedure outlined in Section~\ref{sec:kilo}. These spectra are for a HMNS lifetime of 100 ms with a dynamical torus with $10^{-2}\;M_{\odot}$ of neutron-rich material observed at a viewing angle of $\cos{\theta} = 0.95\; (\theta \approx 18$ deg). For the neutron precursor, we use the {\em r}-band light curve provided by Metzger et al. (2014), for $10^{-4}\;M_{\odot}$ of free neutrons with an electron fraction of $Y_e \approx 0.05$ and opacity of $\kappa \approx 30$ cm$^2$ g$^{-1}$. These light curves are plotted in Figure~\ref{fig:altLC} along with those from the dynamical ejecta-only model for $M_{{\rm ej}} = 10^{-2}\;M_{\odot}$ and $\beta_{{\rm ej}} = 0.2$ used in Sections~\ref{sec:MCsims} and ~\ref{sec:diff}.
   
    \begin{figure*}[t!]
   \centering
	\includegraphics[width=0.7\textwidth]{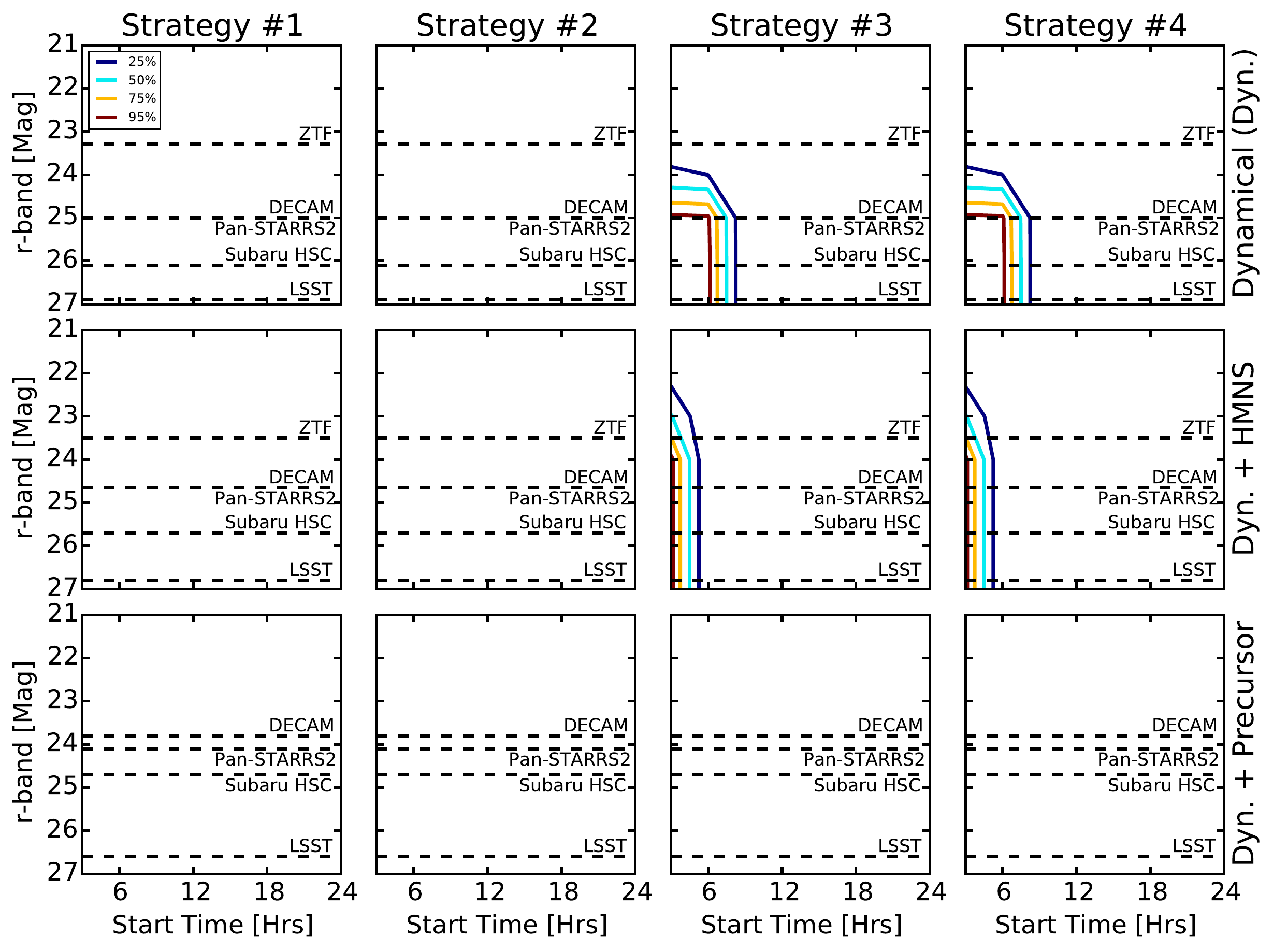}
   \caption{Same as Figure~\ref{fig:altdet}, but showing the fraction of events for which a brightening is detected. Compared to the case of dynamical ejecta only, it is significantly more challenging to detect brightening in these alternative kilonova models due to their earlier peak times and rapid evolution.}
   \label{fig:altrise}
   \end{figure*}
   
We repeat the detectability simulations discussed in Section~\ref{sec:MCsims_det} for the two speculative light curves in Figure~\ref{fig:altLC} which are co-added in flux space to the light curve for the dynamical ejecta from Barnes \& Kasen (2013). The general simulation procedure and criteria for detection are identical to those described in Section~\ref{sec:MCsims_det}. The contours are plotted in Figure~\ref{fig:altdet}. The {\em r}-band contours for the dynamical ejecta only (Figure~\ref{fig:det}) are reproduced for comparison. 

Inspecting the case of a long-lived HMNS, we first note that the typical depth required to detect 50\% (95\%) of events is 23.4 (24) mag across all four observing scenarios. This is shallower than the depth required for the dynamical ejecta alone, but still within reach of the larger telescopes. When a slower cadence strategy is utilized (e.g., Strategy 2 which involves observations every other night after the first observation) the 95\% detection criterion drops to 24.7 mag for observations starting more than six hours after the GW trigger. This is due to the rapid decline of the light curve after peak. Observations must then be deeper at late times to obtain the necessary three epochs for a detection. If a rapid cadence is utilized (e.g., Strategies 3 and 4), then the 50\% detection fraction contour is at 23 mag for observations within six hours of the GW trigger. The other contours are unaffected. 

For the neutron precursor model, the typical depth required to detect 50\% (95\%) of events is 24.7 (25.9) mag for observing Strategies 1 and 2, comparable to those required for the dynamical ejecta alone because the precursor is short-lived. When Strategies 3 and 4 are utilized, the 50\% detection fraction contour begins at 23.6 mag for observations starting 3 hours after the GW trigger. It then monotonically declines to 24.7 mag by 24 hours. This is again because a neutron precursor is short-lived, and only early time observations benefit from its high luminosity.

Contour plots showing the detection of a rise to peak are shown in Figure~\ref{fig:altrise}. As in Section~\ref{sec:MCsims_det}, a rise is defined as any observed brightening between epochs. We note that for Strategies 1 and 2, both of which involve a single observation on the first night, the source is always seen in decline for all three models. In the case of a long-lived HMNS brightening is observed in 50\% (95\%) of events for observations starting $\lesssim5$ ($\lesssim 3)$ hours of the GW trigger and reaching a depth of $\approx$ 23 (24) mag. For a neutron precursor, brightening in the source is never observed. This is because the source peaks in $\sim$ 2 hours, while a reasonable start time is longer.
 
Ultimately, the case of a long-lived HMNS irradiating the ejecta shows the strongest potential impact on kilonova {\em r}-band detectability. However, this component has a strong dependence on viewing angle and thus will not always be observed. In addition, this component is still faint, requiring observations that achieve a depth of $\approx24$ mag in {\em r}-band to detect 95\% of the sources. This scenario is highly speculative and requires further study of the neutron star equation of state and post-merger behavior to assess the likelihood that the post-merger remnant will avoid the prompt collapse to a black hole. The scenario involving a neutron precursor does not drastically impact {\em r}-band detectability unless an observing strategies with rapid intra-night cadence is employed (i.e., a baseline of $\sim1$ hr).

\section{Follow-Up Strategies}
\label{sec:disc}
Motivated by the results of our simulated observations, we now assess the most effective methodologies for conducting efficient GW follow-up observations. There are two primary issues to address: The detection of a kilonovae, and its identification against a population of contaminant sources. Our simulations show that, in the case of a pre-existing template, observations which achieve a depth of $i\approx24$ mag and $z\approx23$ mag lead to the detection of about 95\% of kilonovae. Furthermore, observations should commence within 12 hours of the GW trigger to maximize the likelihood of detecting the source during the rise to peak. We also show that cuts made on $i-z$ color, the rise time (if available), and $\Delta m_z$ are strong discriminants for separating a kilonova from the contaminant population. Quantitatively, with $i-z\gtrsim0$ mag, a rise time of $\lesssim 4$ days, and $|\Delta m_z| \gtrsim 0.4$ mag, we can eliminate a significant fraction $(\gtrsim95\%)$ of contaminants without removing any kilonovae. If the source is seen in decline then requiring $i-z\gtrsim0$ mag and $|\Delta m_z| \gtrsim 0.4$ mag are still strong cuts, eliminating $\sim60\%$ of contaminants. If we also require a decline time of $\lesssim 5$ days then we can eliminate $\gtrsim 90\%$ of contaminants, however this cut eliminates nearly half of the kilonova population. These results do not depend strongly on the choice of cadence (i.e., there is no benefit from intra-night observations). Therefore, follow-up observations should include a single visit per filter per night for at least the first few nights following a trigger.

Taking these constraints into account; when developing the choice of observation strategy there are three questions that must be considered:
\begin{enumerate}
\item How large of an area must be covered by the observations?
\item How long is the region of interest observable? 
\item How soon after the GW trigger can observations commence?
\end{enumerate}

\begin{deluxetable*}{cccccccccc}
\tabletypesize{\footnotesize}
\tablecolumns{9} 
\tabcolsep0.05in\footnotesize
\tablewidth{0pt}  
\tablecaption{Follow-up Capabilities of Wide-Field Telescopes}
\tablehead{\multicolumn{3}{c}{} & \multicolumn{2}{c}{{\em r}-band} & \multicolumn{2}{c}{{\em i}-band} & \multicolumn{2}{c}{{\em z}-band} & \multicolumn{1}{c}{} \\
\colhead{Instrument} & \colhead{FoV} & \colhead{Etendue} & \colhead{Exp.} & \colhead{Rate}  & \colhead{Exp.}  &\colhead{Rate} & \colhead{Exp.} & \colhead{Rate} &  \colhead{$i + z$ Rate} \\
\colhead{} & \colhead{(deg$^2$)} & \colhead{(m$^2$ deg$^2$)} & \colhead{(s)} & \colhead{(deg$^2$ hr$^{-1}$)} & \colhead{(s)} & \colhead{(deg$^2$ hr$^{-1}$)} & \colhead{(s)} & \colhead{(deg$^2$ hr$^{-1}$)} &  \colhead{(deg$^2$ hr$^{-1}$)}}
\startdata
  DECam$^{{\rm a}}$ & 3 &  37.7 & 4200 & 2.5 & 250 & 43 & 150 & 72 & 27  \\
 ZTF$^{{\rm b}}$ & 47 & 53.2 & $9\times10^5$  & 0.2 & 1.6$\times10^4$ & 10 & {\bf --} & {\bf --}  & {\bf --}   \\
 Subaru HSC$^{{\rm c}}$  & 1.8  & 90.5 & 520 & 12 & 55 & 115 & 45 & 145 & 65 \\
   Pan-STARRS2$^{{\rm d}}$ & 14 & 35.6 & $1.4\times10^4$  & 4 & 700  & 72 & 350 & 144  & 30 \\
    LSST$^{{\rm e}}$ & 9.6 & 315 & 390 & 88 & 28 & 1235  & 19 & 1820 & 735
 \enddata
\tablecomments{Expected rates of sky coverage in deg$^2$ hr$^{-1}$ for several existing and future wide-field facilities at a depth of $r\approx26$ mag, $i\approx24$ mag and $z\approx23$ mag (i.e., the depths necessary to achieve a 95\% detection rate for kilonovae). We note that for HSC and LSST we include overhead times of 20s and 2s, respectively, as they are comparable to the exposure times. 
References: \begin{enumerate}[label=\alph*,topsep=5pt]
\item DECam ETC, \url{http://www.ctio.noao.edu/noao/content/Exposure-Time-Calculator-ETC-0}
\item Nissanke et al.(2013), Bellm (2014), Kasliwal \& Nissanke (2014)
\item HSC ETC, \url{http://www.naoj.org/cgi-bin/img_etc.cgi}
\item Pan-STARRS DRM, \url{http://pan-starrs.ifa.hawaii.edu/}
\item LSST ETC, \url{http://dls.physics.ucdavis.edu:8080/etc4_3work/servlets/LsstEtc.html}
\end{enumerate}
}
\label{tab:tele}
\end{deluxetable*}

The first and second points are the most important considerations as the sizable localization regions from the early science-runs of Advanced LIGO present the largest observational challenge and have the most pronounced impact on the quality of data that can be obtained. The key issue here is whether or not there will be sufficient time to observe the entire GW localization region on a nightly basis. A successful GW follow-up campaign must be able to observe a significant fraction of the GW localization region in a single night to obtain the optimal cadence. The third question impacts the likelihood that we will be able to observe a kilonovae before peak. This is important as we have shown in Sections~\ref{sec:MCsims_cont} and~\ref{sec:diff} that it is easier to isolate a kilonova from the contaminant population if it is observed during the rise. Additionally, observing a kilonova at early times is important both for constraining the ejecta parameters and for testing the alternative kilonovae models outlined in Section~\ref{sec:altkilo}. Ultimately, our ability to address the three questions above depends heavily on the facilities in use.

To investigate the impact of the choice of observing facility on GW follow-up we determine the expected performance of several current and future wide-field telescopes. Specifically, we consider the performance of DECam, ZTF, Subaru Hyper Suprime-Cam (HSC), Pan-STARRS2, and LSST. We take into account the field-of-view (FoV) per pointing, the required integration time to achieve a depth of $r\approx26$, $i\approx24$, and $z\approx23$ mag, and consequently how much area can be covered per hour of observing time. We also compute the sky coverage rate when observing in $i$- and $z$-band to obtain color data. These values are given in Table~\ref{tab:tele}. 

We first look to currently available facilities (e.g., DECam, HSC, and Pan-STARRS2). In the Northern hemisphere the two available telescopes are Pan-STARRS2 and HSC. Pan-STARRS2 offers an impressive FoV ($\sim14$ deg$^2$) allowing a $\sim100$ deg$^2$ region to be imaged in just $8$ pointings. However, HSC is the clear front-runner as  the immense collecting area allows observations to achieve the target depths in much shorter integration times than Pan-STARRS2. Thus, despite the smaller FoV of HSC, the sky coverage rate for $i$- and $z$-band data is faster than Pan-STARRS2 by over a factor of two. If we consider a typical GW localization region ($\sim 100$ deg$^2$) then Pan-STARRS2 can image this region to the required $i$- and $z$-band depths in $\sim 3.5$ hours while achieving this goal with HSC will require only $\sim1.5$ hours. In the Southern hemisphere DECam is the ideal choice. The wide FoV and red-sensitive CCD of DECam allow it to quickly image a typical GW localization region. If we again consider a $\sim100$ deg$^2$  localization region, then DECam can obtain $i$- and $z$-band data at the required depths in $\sim3.5$ hours. 

We now consider future wide-field telescopes. The next major instrument to become available will be ZTF. The primary advantage of ZTF is the extremely wide FoV $(\sim47$ deg$^2$), allowing even large localization regions to be imaged in just a handful of pointings. However, the longer integration times required to achieve the target depths means that the sky coverage rate of ZTF will be slower than DECam, Pan-STARRS2, and HSC in $r$- and $i$-band. Furthermore, the lack of $z$-band capabilities makes obtaining $i-z$ color data impossible (observations in $i$-band alone may still be useful, see below). Looking further ahead, GW follow-up with LSST will be particularly effective. The large collecting area and high sensitivity allow LSST to image a GW localization region with unparalleled rapidity. If we again consider a typical localization region ($\sim 100$ deg$^2$), then ZTF will require $\sim500$ hours in $r$-band and and $\sim10$ hours in $i$-band to image the entire region to the required depths. By comparison, LSST can image the entire localization region to the required depths in $i$- and $z$-band in just $\sim10$ minutes, highlighting the immense observing power of future facilities. However, LSST is not expected to begin science operations until $\sim2022$ when GW localizations will be much improved.

While GW follow-up observations should strive to obtain $i-z$ color data as outlined above, this may not be possible because either there is insufficient time to observe the localization region in $i+z$ or because $z$-band observations are not feasible (e.g., ZTF). In this scenario, data may be taken in a single filter,  the identification of a kilonova becomes difficult. We dedicate the rest of this section to a brief discussion of eliminating contaminants when only single filter observations are possible. 

The most populous contaminant to contend with is the Type Ia SNe. In the context of observations utilizing a single filter, these contaminants are most easily eliminated by utilizing their much slower rise and decline timescales. Therefore, the best method to reject these sources is to perform image subtraction using observations that are separated by (e.g., $\gtrsim1$ day between epochs). Type Ia SNe are unlikely to be detected in these subtractions as they are unable to produce a statistically significant change in flux between the two epochs. Qualitatively, if we require a source to exhibit a change in flux of at least $|\Delta m_z| \gtrsim 0.4$ mag and a rise time of $\lesssim 4$ days, then we can eliminate $99\%$ of detected Type Ia SNe. This corresponds to $\sim6$ detected events remaining after the cuts have been applied. However, we also eliminate $\sim15-35\%$ of detected kilonova, with the higher fraction of sources eliminated in Strategies 3 and 4 where the rapid cadence of observations on the first night produces smaller values of $|\Delta m_z|$. For sources seen on the decline, if we require that the decline time is $\lesssim 5$ days and the change in flux is $|\Delta m_z| \gtrsim 0.4$, then we eliminate all of the detected Type Ia SNe, while rejecting only $\sim2\%$ of detected kilonovae. In both scenarios (e.g., sources seen to rise or only in decline) rejected kilonovae can be recovered using a late-time template image, but real time detection may be difficult. Still, despite their high expected rates, Type Ia SNe will likely prove to be one of the easiest contaminants to eliminate in GW follow-up searches. 

We can also apply the above argument to Pan-STARRS fast transients that are seen only in decline. Qualitatively, if we require that sources exhibit a decline time of $\lesssim 5$ days and a change in flux of $|\Delta m_z| \gtrsim 0.4$, then we can eliminate $\sim80\%$ of the detected Pan-STARRS fast transients, again while rejected $\sim2\%$ of the detected kilonovae. While the cuts on decline time and $|\Delta m_z|$ are not as effective here as in the case of Type Ia SNe, the lower rates ($\sim 7$ total expected during a typical GW follow-up search) means that we expect only $\sim2$ detected events to survive. These remaining sources can then be followed-up and eliminated by utilizing pointed observations as described below.

While inter-night variability is an effective tool for eliminating slowly evolving contaminants, it is not feasible for the rapidly-evolving contaminants (e.g., type .Ia, miscellaneous fast transients, and Pan-STARRS fast transients observed on the rise) as they exhibit timescales and changes in brightness comparable to those of kilonovae. We can rule out type .Ia and miscellaneous fast transients as likely contaminants by utilizing the rate information from Section~\ref{sec:rates}. We showed that the rates for such events are low, resulting in $\lesssim1$ source expected on the timescale and in the area of a typical GW follow-up campaign. We further showed in Section~\ref{sec:diff}, that if the rate information is convolved with the detection rates in our simulations, then we expect $\ll 1$ of these contaminants to appear during the course of follow-up observations. In this case, even considering the fact that the rates are poorly constrained, it is unlikely that these sources will lead to significant contamination. However, this rate argument does not apply to the Pan-STARRS fast transients as the rates of these events are sufficiently high ($\sim 7$ expected during follow-up of a typical GW error region) that they are likely to represent significant contamination. In this case, the best option is to include any such sources in a list of kilonova candidates for follow up with pointed observations as described below or wait until the source begins to decline and apply the inter-night variability arguments outlined above. 

Lastly, we consider flaring stars (e.g., M-Dwarfs, Section~\ref{sec:flarestar}) as such events are likely to become contaminants in observing strategies that rely on a single epoch to identify sources. This is particularly true for observations carried out in bluer bands (e.g., $g$ or $r$) where the flares amplitudes are larger and the kilonova timescales are much shorter. Given that flares generally exhibit short timescales (minutes to hours, Berger et al. 2013) then if we require that a source be detected in at least two epochs separated by $\gtrsim1$ day we can efficiently eliminate all flare contaminants. The primary concern here is that rapidly evolving kilonovae (e.g., in $r$ band around peak) will also be eliminated, particularly if observations are unable to achieve a depth of $r\approx26$ mag resulting in the kilonovae only being detected in one or two epochs. Additionally, in $r$-band, the quiescent counterpart may be difficult to identify in the template images. These complications can be avoided by carrying out follow-up observations in $i$- or $z$-band.

Taking the above considerations into account, we conclude that single filter observations, particularly those conducted in $z$-band, are still viable for kilonova identification. We favor data taken in $z$-band as the shallower depths required to achieve a 95\% detection rate (when compared to the depths required to achieve a comparable detection rate in $r$- or $i$-band) allow substantially more sky area to be covered in a night of observations.  Furthermore, we have shown above that observations in $z$-band still allow robust contaminant rejection (although still less robust than when color data is available) without being susceptible to additional contamination from stellar flares as may be the case for observations in $r$- and $i$-band. We note that observational strategies that utilize a single filter are not likely to  produce a single, obvious kilonova. This may be particularly true when a pre-existing template is unavailable, reducing the efficacy of our cuts on the timescale and change in brightness. However, the strategies outlined in this section can still be utilized to generate a list of kilonova candidates. In this scenario, once a list of $\lesssim10$ candidates is generated, those sources can be further observed in the NIR and spectroscopically. NIR follow-up is crucial as kilonova are substantially redder than the other contaminants considered, with the kilonova spectrum peaking at $\sim1.5\,\mu$m. Furthermore, spectroscopic observations will allow us to look for features consistent with the production of $r$-process elements as another means of kilonova identification. These observations will vastly improve the robustness of kilonova identification and will allow us to maximize the science gained from a GW event. 

The worst-case scenario is a GW localization region that is simply too large to observe at the required depth in the available time, even in a single filter. Here, there are two possible approaches. The first is to simply obtain shallower observations at a depth that allows the entire GW localization region to be imaged. This will be useful if the source is located closer than the fiducial distance of 200 Mpc considered here, or if the ejecta mass is larger than the fiducial value of 0.01 $M_{\odot}$. However, this is risky as shallower observations are unlikely to detect the typical event. Furthermore, even if a kilonova is detected, it may only appear in one or two epochs, making it more difficult to distinguish from stellar flares. A less risky approach is to carry out the observations at the required depth, but cover the highest probability portion of the localization region, thereby maximizing the likelihood of detecting the kilonova. 

\section{Conclusions}
\label{sec:conc}
The era of gravitational wave astronomy is fast approaching, with ground-based GW interferometers expected to achieve direct detections of the inspiral and merger of compact object binaries sometime during the next few years. If we wish the maximize the science gains of these detections then the rapid identification of an EM counterpart is essential.  Observational and theoretical evidence suggests a host of potential EM counterparts spanning a wide range of timescales, luminosities, and wavelengths. Among these, the isotropic optical emission powered by the radioactive decay of $r$-process elements synthesized in the merger ejecta has proven to be a particularly enticing target. A major challenge for follow-up observations is the large sky area that needs to be observed $(\sim100\text{ deg}^2)$ combined with a substantial depth to detect a typical event. Consequently, optical searches for kilonovae will be plagued by contamination from other common transient events (e.g., Type Ia SNe) as well as from other rapidly evolving transients (e.g., see Section~\ref{sec:contaminants}). We have performed comprehensive simulations designed to probe the range of depths and cadences that may be involved in a GW follow-up campaign to determine the set of observations that will allow kilonove detections coupled with efficient rejection of contaminants. We find that:

\begin{itemize}
\item In the case where a pre-existing template image of the target region is available (Section~\ref{sec:MCsims_det}), the optimal strategy for kilonova detection is to image the region to $i\approx 24$ and $z\approx 23$ mag with a single visit per night. We have shown that these limits are necessary to achieve a 95\% kilonova detection rate. Furthermore, observations should ideally commence within 12 hours of the GW trigger to facilitate the detection of a rise to peak in the light curve. 
\item For sources observed on the rise, cuts on $i-z$ color and rise time are effective at distinguishing kilonovae from the contaminant population (Section~\ref{sec:MCsims_cont}). Specifically, with $i-z\gtrsim0$ mag and a rise time of $\lesssim 4$ days, it is possible to eliminate $\gtrsim95\%$ of contaminants without rejecting any kilonovae. Applying the same cuts to sources seen only in decline is less effective. If we require $i-z\gtrsim0$ mag and a decline time of $\lesssim5$ days then we again eliminate $\gtrsim95\%$ of the contaminant population, but lose nearly half of the kilonovae. These rejected kilonovae can be recovered using a late-time template $(\gtrsim10$ days), but real-time kilonova identification will be hampered.
\item In the case where a pre-existing template is not available (Section~\ref{sec:diff}), we find that there is no effect on the optimal cadence, but observations must go $\sim1$ magnitude deeper than in the pre-existing template case to achieve comparable detection rates. The detection of a rise is also more difficult since the first observation must be used as a template image, reducing the amount of early-time data.
\item In the case where a pre-existing template template is not available, cuts on $i-z$ color and timescale are still effective for separating kilonovae from the contaminant population for sources observed during the rise to peak (Section~\ref{sec:diff}). However, for sources seen only in decline, these cuts are significantly less effective. We do note that by utilizing a detection criterion based on the change in brightness in an observed source we can reduce the number of contaminant sources detected. However, this also reduces the kilonovae detection rates. 
\item For the alternative kilonova models (Section~\ref{sec:altkilo}) we find that $r$-band detectability in the speculative case of a long-lives HMNS producing an early-time blue component may improve, but the short-lived nature of this component combined with the strong dependance on viewing angle means that it is not likely to be observed. If the kilonova is accompanied by a neutron precursor (Section~\ref{sec:neutronpre}), then this component will appear brighter than a kilonova ($r\approx22.2$ mag), but the extremely short timescale ($\lesssim2$ hrs) makes it unlikely to appear in follow-up observations.
\item We examine the potential of current and future wide-field telescope facilities to achieve the observational goals outlined above (Section~\ref{sec:disc}). For current facilities, we find that Subaru HSC is the best option for localization regions in the Northern hemisphere while DECam is best suited to GW follow-up in the Southern hemisphere. Both telescopes can easily image an entire GW localization region to the required depths in $i$- and $z$-band at the required cadence. In the future, GW follow-up with LSST will be particularly effective and rapid.
\item  If data in both $i$- and $z$-band cannot be obtained due to observational constraints, then an attempt at kilonova identification can be made using data from a single filter, ideally $z$-band (Section~\ref{sec:disc}). We found that without color data it is still possible to eliminate the most numerous contaminants (e.g., Type Ia SNe and Pan-STARRS fast transients) by utilizing their slow light curve evolution when compared to kilonovae. The rapidly evolving contaminants (e.g., type .Ia and miscellaneous fast transients) are not likely to be a significant contaminant on the basis of their low expected rates. Lastly, M-dwarf stellar flares are easily rejected by requiring that the source appear in at least two epochs or by identifying a quiescent counterpart in the template image.
\end{itemize}

The next step of this analysis is to test the methodologies and conclusions of this work against a set of real data. As discussed in Section~\ref{sec:intro}, we have obtained DECam $i$- and $z$-band data with the cadence and depth optimized for GW follow-up over an area of $\sim$ 70 deg$^2$. We will confront the results of this paper by using subsets of these data to mimic the four observing strategies as well as the effects of having or missing a pre-existing template (utilizing follow-up observations that we obtained about two weeks after the main survey). This will facilitate a better observational understanding of the {\em true} contaminant population on timescales relevant to GW follow-up and further refine any proposed strategies.

\vspace{5pt}
We thank Ryan Chornock, Maria Drout, Wen-fai Fong, Ryan Foley, Daniel Kasen, Brian Metzger, Armin Rest, and Ken Shen for helpful discussions and providing model data during the course of this analysis. P.S.C. is grateful for support provided by the NSF through the Graduate Research Fellowship Program, grant DGE1144152.


\begin{thebibliography}{}
\bibitem[Aasi et al.(2014)]{2014ApJS..211....7A} Aasi, J., Abadie, J., 
Abbott, B.~P., et al.\ 2014, \apjs, 211, 7
\bibitem[Abadie et al.(2010)]{2010CQGra..27q3001A} Abadie, J., Abbott, 
B.~P., Abbott, R., et al.\ 2010, Classical and Quantum Gravity, 27, 173001
\bibitem[Abbott et al.(2009)]{2009RPPh...72g6901A} Abbott, B.~P., Abbott, 
R., Adhikari, R., et al.\ 2009, Reports on Progress in Physics, 72, 076901
\bibitem[Acernese et al.(2009)]{2009CQGra..26h5009A} Acernese, F., 
Alshourbagy, M., Antonucci, F., et al.\ 2009, Classical and Quantum 
Gravity, 26, 085009
\bibitem[Barnes 
\& Kasen(2013)]{2013ApJ...775...18B} Barnes, J., \& Kasen, D.\ 2013, \apj, 775, 18
\bibitem[Bellm(2014)]{2014htu..conf...27B} Bellm, E.\ 2014, The Third 
Hot-wiring the Transient Universe Workshop, 27 
\bibitem[Berger(2011)]{2011NewAR..55....1B} Berger, E.\ 2011, \nar, 55, 1
\bibitem[Berger et al.(2013)]{2013ApJ...779...18B} Berger, E., Leibler, 
C.~N., Chornock, R., et al.\ 2013, \apj, 779, 18 
\bibitem[Berger(2014)]{2014ARA&A..52...43B} Berger, E.\ 2014, \araa, 52, 43
\bibitem[Bildsten et al.(2007)]{2007ApJ...662L..95B} Bildsten, L., Shen, 
K.~J., Weinberg, N.~N., \& Nelemans, G.\ 2007, \apjl, 662, L95
\bibitem[Bloom et al.(2006)]{2006ApJ...638..354B} Bloom, J.~S., Prochaska, 
J.~X., Pooley, D., et al.\ 2006, \apj, 638, 354 v
\bibitem[Bauswein et al.(2013a)]{2013PhRvL.111m1101B} Bauswein, A., 
Baumgarte, T.~W., \& Janka, H.-T.\ 2013a, Physical Review Letters, 111, 131101
\bibitem[Bauswein et al.(2013b)]{2013ApJ...773...78B} Bauswein, A., Goriely, 
S., \& Janka, H.-T.\ 2013b, \apj, 773, 78
\bibitem[Chandra 
\& Frail(2012)]{2012ApJ...746..156C} Chandra, P., \& Frail, D.~A.\ 2012, \apj, 746, 156
\bibitem[Darbha et al.(2010)]{2010MNRAS.409..846D} Darbha, S., Metzger, 
B.~D., Quataert, E., et al.\ 2010, \mnras, 409, 846
\bibitem[Dessart et al.(2009)]{2009ApJ...690.1681D} Dessart, L., Ott, 
C.~D., Burrows, A., Rosswog, S., \& Livne, E.\ 2009, \apj, 690, 1681
\bibitem[Drout et al.(2013)]{2013ApJ...774...58D} Drout, M.~R., Soderberg, 
A.~M., Mazzali, P.~A., et al.\ 2013, \apj, 774, 58
\bibitem[Drout et al.(2014)]{2014ApJ...794...23D} Drout, M.~R., Chornock, 
R., Soderberg, A.~M., et al.\ 2014, \apj, 794, 23
\bibitem[Fairhurst(2009)]{2009NJPh...11l3006F} Fairhurst, S.\ 2009, New 
Journal of Physics, 11, 123006
\bibitem[Fern{\'a}ndez 
\& Metzger(2013)]{2013MNRAS.435..502F} Fern{\'a}ndez, R., \& Metzger, B.~D.\ 2013, \mnras, 435, 502
\bibitem[Fern{\'a}ndez et al.(2014)]{2014arXiv1409.4426F} Fern{\'a}ndez, 
R., Kasen, D., Metzger, B.~D., \& Quataert, E.\ 2014, arXiv:1409.4426
\bibitem[Fong et al.(2010)]{2010ApJ...708....9F} Fong, W., Berger, E., 
\& Fox, D.~B.\ 2010, \apj, 708, 9
\bibitem[Fong et al.(2011)]{2011ApJ...730...26F} Fong, W., Berger, E., 
Chornock, R., et al.\ 2011, \apj, 730, 26
\bibitem[Fong 
\& Berger(2013)]{2013ApJ...776...18F} Fong, W., \& Berger, E.\ 2013, \apj, 776, 18
\bibitem[Fong et al.(2013)]{2013ApJ...769...56F} Fong, W., Berger, E., 
Chornock, R., et al.\ 2013, \apj, 769, 56
\bibitem[Fryer et al.(1999)]{1999ApJ...520..650F} Fryer, C.~L., Woosley, 
S.~E., Herant, M., \& Davies, M.~B.\ 1999, \apj, 520, 650 
\bibitem[Graur et al.(2014)]{2014ApJ...783...28G} Graur, O., Rodney, S.~A., 
Maoz, D., et al.\ 2014, \apj, 783, 28 
\bibitem[Goriely et al.(2013)]{2013PhRvL.111x2502G} Goriely, S., Sida, 
J.-L., Lema{\^i}tre, J.-F., et al.\ 2013, Physical Review Letters, 111, 
242502 
\bibitem[Grossman et al.(2014)]{2014MNRAS.439..757G} Grossman, D., 
Korobkin, O., Rosswog, S., \& Piran, T.\ 2014, \mnras, 439, 757
\bibitem[Harry 
\& Fairhurst(2011)]{2011PhRvD..83h4002H} Harry, I.~W., \& Fairhurst, S.\ 2011, \prd, 83, 08400 
\bibitem[Hughes 
\& Holz(2003)]{2003CQGra..20S..65H} Hughes, S.~A., \& Holz, D.~E.\ 2003, Classical and Quantum Gravity, 20, 65
\bibitem[Just et al.(2014)]{2014arXiv1406.2687J} Just, O., Bauswein, A., 
Ardevol Pulpillo, R., Goriely, S., \& Janka, H.-T.\ 2014, arXiv:1406.2687
\bibitem[Kaplan et al.(2014)]{2014ApJ...790...19K} Kaplan, J.~D., Ott, 
C.~D., O'Connor, E.~P., et al.\ 2014, \apj, 790, 19
\bibitem[Kasen et al.(2013)]{2013ApJ...774...25K} Kasen, D., Badnell, 
N.~R., \& Barnes, J.\ 2013, \apj, 774, 25
\bibitem[Kasliwal 
\& Nissanke(2014)]{2014ApJ...789L...5K} Kasliwal, M.~M., \& Nissanke, S.\ 2014, \apjl, 789, LL5 
\bibitem[Li \& Paczy{\'n}ski(1998)]{1998ApJ...507L..59L} Li, L.-X., \& Paczy{\'n}ski, B.\ 1998, \apjl, 507, L59
\bibitem[Li et al.(2011)]{2011MNRAS.412.1473L} Li, W., Chornock, R., 
Leaman, J., et al.\ 2011, \mnras, 412, 1473
\bibitem[Metzger et al.(2009a)]{2009MNRAS.396..304M} Metzger, B.~D., Piro, 
A.~L., \& Quataert, E.\ 2009a, \mnras, 396, 304 
\bibitem[Metzger et al.(2009b)]{2009MNRAS.396.1659M} Metzger, B.~D., Piro, 
A.~L., \& Quataert, E.\ 2009b, \mnras, 396, 1659
\bibitem[Metzger et al.(2010)]{2010MNRAS.406.2650M} Metzger, B.~D., 
Mart{\'{\i}}nez-Pinedo, G., Darbha, S., et al.\ 2010, \mnras, 406, 2650
\bibitem[Metzger 
\& Berger(2012)]{2012ApJ...746...48M} Metzger, B.~D., \& Berger, E.\ 2012, \apj, 746, 48
\bibitem[Metzger(2012)]{2012MNRAS.419..827M} Metzger, B.~D.\ 2012, \mnras, 
419, 827
\bibitem[Metzger 
\& Piro(2014)]{2014MNRAS.439.3916M} Metzger, B.~D., \& Piro, A.~L.\ 2014, \mnras, 439, 3916
\bibitem[Metzger et al.(2014)]{2014arXiv1409.0544M} Metzger, B.~D., 
Bauswein, A., Goriely, S., \& Kasen, D.\ 2014, arXiv:1409.0544
\bibitem[Metzger 
\& Fern{\'a}ndez(2014)]{2014MNRAS.441.3444M} Metzger, B.~D., \& Fern{\'a}ndez, R.\ 2014, \mnras, 441, 3444
\bibitem[Nakar 
\& Piran(2011)]{2011Natur.478...82N} Nakar, E., \& Piran, T.\ 2011, \nat, 478, 82
\bibitem[Narayan et al.(1992)]{1992ApJ...395L..83N} Narayan, R., Paczynski, 
B., \& Piran, T.\ 1992, \apjl, 395, L83
\bibitem[Nissanke et al.(2011)]{2011ApJ...739...99N} Nissanke, S., Sievers, 
J., Dalal, N., \& Holz, D.\ 2011, \apj, 739, 99
\bibitem[Nissanke et al.(2013)]{2013ApJ...767..124N} Nissanke, S., 
Kasliwal, M., \& Georgieva, A.\ 2013, \apj, 767, 124
\bibitem[Nugent et al.(2011a)]{2011ATel.3581....1N} Nugent, P., Sullivan, 
M., Bersier, D., et al.\ 2011a, The Astronomer's Telegram, 3581, 1
\bibitem[Nugent et al.(2011)]{2011Natur.480..344N} Nugent, P.~E., Sullivan, 
M., Cenko, S.~B., et al.\ 2011, \nat, 480, 344 
\bibitem[Paczynski(1986)]{1986ApJ...308L..43P} Paczynski, B.\ 1986, \apjl, 
308, L43 
\bibitem[Pereira et 
al.(2013)]{2013A&A...554A..27P} Pereira, R., Thomas, R.~C., Aldering, G., et al.\ 2013, \aap, 554, AA27
\bibitem[Perrett et al.(2012)]{2012AJ....144...59P} Perrett, K., Sullivan, 
M., Conley, A., et al.\ 2012, \aj, 144, 59 
\bibitem[Phinney(2009)]{2009astro2010S.235P} Phinney, E.~S.\ 2009, 
astro2010: The Astronomy and Astrophysics Decadal Survey, 2010, 235
\bibitem[Planck Collaboration et 
al.(2014)]{2014A&A...571A..16P} Planck Collaboration, Ade, P.~A.~R., Aghanim, N., et al.\ 2014, \aap, 571, AA16
\bibitem[Rest et al.(2014)]{2014ApJ...795...44R} Rest, A., Scolnic, D., 
Foley, R.~J., et al.\ 2014, \apj, 795, 44
\bibitem[Rosswog et 
al.(1999)]{1999A&A...341..499R} Rosswog, S., Liebend{\"o}rfer, M., Thielemann, F.-K., et al.\ 1999, \aap, 341, 499
\bibitem[Rosswog(2005)]{2005ApJ...634.1202R} Rosswog, S.\ 2005, \apj, 634, 
1202 
\bibitem[Rosswog et al.(2013)]{2013MNRAS.430.2585R} Rosswog, S., Piran, T., 
\& Nakar, E.\ 2013, \mnras, 430, 2585 
\bibitem[Schlafly 
\& Finkbeiner(2011)]{2011ApJ...737..103S} Schlafly, E.~F., \& Finkbeiner, D.~P.\ 2011, \apj, 737, 103
\bibitem[Sekiguchi et al.(2011)]{2011PhRvL.107e1102S} Sekiguchi, Y., 
Kiuchi, K., Kyutoku, K., 
\& Shibata, M.\ 2011, Physical Review Letters, 107, 051102
\bibitem[Shappee 
\& Stanek(2011)]{2011ApJ...733..124S} Shappee, B.~J., \& Stanek, K.~Z.\ 2011, \apj, 733, 124
\bibitem[Shen et al.(2010)]{2010ApJ...715..767S} Shen, K.~J., Kasen, D., 
Weinberg, N.~N., Bildsten, L., \& Scannapieco, E.\ 2010, \apj, 715, 767
\bibitem[Sim et al.(2012)]{2012MNRAS.420.3003S} Sim, S.~A., Fink, M., 
Kromer, M., et al.\ 2012, \mnras, 420, 3003
\bibitem[Stamatikos et al.(2009)]{2009astro2010S.284S} Stamatikos, M., 
Gehrels, N., Halzen, F., M{\'e}sz{\'a}ros, P., 
\& Roming, P.~W.~A.\ 2009, astro2010: The Astronomy and Astrophysics Decadal Survey, 2010, 284 
\bibitem[Stubbs(2008)]{2008CQGra..25r4033S} Stubbs, C.~W.\ 2008, Classical 
and Quantum Gravity, 25, 184033 
\bibitem[Sylvestre(2003)]{2003ApJ...591.1152S} Sylvestre, J.\ 2003, \apj, 
591, 1152 
\bibitem[Tanaka 
\& Hotokezaka(2013)]{2013ApJ...775..113T} Tanaka, M., \& Hotokezaka, K.\ 2013, \apj, 775, 113
\bibitem[Tanaka et al.(2014)]{2014ApJ...780...31T} Tanaka, M., Hotokezaka, 
K., Kyutoku, K., et al.\ 2014, \apj, 780, 31 
\bibitem[Thompson et al.(2009)]{2009arXiv0912.0009T} Thompson, T.~A., 
Kistler, M.~D., \& Stanek, K.~Z.\ 2009, arXiv:0912.0009
\bibitem[van Eerten et al.(2010)]{2010ApJ...722..235V} van Eerten, H., 
Zhang, W., \& MacFadyen, A.\ 2010, \apj, 722, 235
\bibitem[van Eerten 
\& MacFadyen(2011)]{2011ApJ...733L..37V} van Eerten, H.~J., \& MacFadyen, A.~I.\ 2011, \apjl, 733, LL37
\end{thebibliography}
\end{document}